\documentclass[superscriptaddress,nofootinbib]{revtex4-1}
\usepackage{dcolumn}
\usepackage[utf8]{inputenc}
\usepackage{amsmath}
\usepackage{amsfonts}
\usepackage{amssymb}
 \usepackage{hyperref}
\usepackage{graphicx}
\usepackage{hyperref}
\usepackage{xcolor}
\newcommand{\CNOT}{\textsc{cnot}}

\newcommand{\ket}[1]{\ensuremath{|{#1}\rangle}}

\usepackage{tikz}
\usetikzlibrary{quantikz}
\begin{document}

\title{Quantum-information methods for quantum gravity laboratory-based tests}

\author{Chiara Marletto, Vlatko Vedral}
\affiliation{Clarendon Laboratory, University of Oxford, Parks Road, Oxford OX1 3PU, United Kingdom}

\date{\today}%

\begin{abstract}
Quantum theory and general relativity are about one century old. At present, they are considered the best available explanations of physical reality, and they have been so far corroborated by all experiments realised so far. Nonetheless, the quest to unify them is still ongoing, with several yet untested proposals for a theory of quantum gravity. Here we review the nascent field of information-theoretic methods applied to designing tests of quantum gravity in the laboratory. This field emerges from the fruitful extension of quantum information theory methodologies beyond the domain of applicability of quantum theory itself, to cover gravity. {We shall focus mainly on the detection of gravitational entanglement between two quantum probes, comparing this method with single-probe schemes}. We shall review the experimental proposal that has originated this field, as well as its variants, their applications, and discuss their potential implications for the quantum theory of gravity. We shall also highlight the role of general information-theoretic principles in illuminating the search for quantum effects in gravity.  
\end{abstract}

\maketitle

\tableofcontents

\section{Introduction}

Quantum theory and general relativity are the best existing explanations of physical reality and they have so far proven unchallenged by experimental evidence. Yet, they are fundamentally incompatible. Quantum theory allows for phenomena such as quantum superpositions, while general relativity does not: it is a classical theory, {that does not allow for quantum effects - just like Newton's laws of gravitation}. Numerous arguments have been proposed supporting the idea that quantum theory ceases to apply at a certain scale, for instance due to the collapse of macroscopic quantum superpositions, \cite{GHIBAS}.  Gravity is often considered as the culprit for this phenomenon. On the other hand many proposals for a quantum theory of gravity have been put forward, \cite{KIE}. {However, no definitive experimental confirmation of any quantum gravity theory has so far been achieved. In fact, until very recently, it was thought that experimental tests of any quantum gravity effect are far beyond current experimental capabilities, \cite{DYS}}.  From time to time, it has even been claimed that quantising gravity is not necessary because (the argument goes) it leads to untestable consequences, \cite{ROBO06}. {Hereinafter we will use the terms ``quantum theory of gravity" and ``quantum gravity" to describe any physical theory that puts together general relativity and quantum theory, and is applicable to all regimes (both low-energy and high-energy regimes). We note this differs from the terminology in some of the existing literature where `quantum gravity' is sometimes reserved exclusively for the high-energy quantum theories of gravity.} 

Most proposals to test quantum gravity are not within reach of present-day technology. One proposal to probe quantum gravity is through high-energy particle colliders, such as the Large Hadron Collider, \cite{HIGH}. These colliders can create extremely energetic conditions similar to those that existed shortly after the Big Bang. By studying the high-energy collisions of particles, it is hoped to observe potential deviations from the predictions of classical general relativity that could hint at the presence of quantum gravitational effects. 
Another potential avenue is provided by gravitational waves. Advanced detectors such as LIGO (Laser Interferometer Gravitational-Wave Observatory) have already detected gravitational waves, confirming some of the predictions of general relativity, \cite{LIGO}. {On a more speculative note, future gravitational wave observatories, such as the proposed Einstein telescope, could in principle observe subtle deviations in the properties of gravitational waves that would indicate quantum gravitational effects, see e.g. \cite{ADD5}.} {Another possible path to link quantum effects to laboratory tests is analogue gravity, \cite{ANA}, which aims at simulating semiclassical gravity effects via quantum systems in the lab, such as cold atoms, superfluids, superconducting qubits, optical systems.}
More exotic approaches include probing the physics of black holes, \cite{BLA}. {Studying the properties of black holes could provide insights into the quantum nature of gravity. For instance, some believe that detecting signatures of quantum coherence during black hole evaporation or testing theoretical predictions related to the information paradox would be valuable tests for quantum gravity; others maintain that properties of black holes have yet to be properly understood and find this avenue less promising,\cite{WALD}.} Finally, the study of the early universe and its large-scale structures could provide valuable information about the nature of gravity at high energies,\cite{WIL}. Cosmological observations, such as the cosmic microwave background radiation or the large-scale distribution of galaxies, may also provide evidence of primordial gravitational waves or other phenomena that would have originated during the quantum gravity era. {A possible avenue is also to use experiments involving high-precision measurements of fundamental constants, such as the gravitational constant or the fine structure constant; these may reveal variations or deviations from their expected values. Such variations could suggest underlying quantum gravitational effects at play, see for instance the review in \cite{KIE2}. However, the experimental verification of specific quantum gravity theories via the above methods remains a challenge, and it requires huge advancements in technology and novel experimental approaches.}

In this review, we shall discuss a recently proposed approach to testing quantum effects in gravity, that overcomes these difficulties and is based on information-theoretic methods. It promises to be a game-changer, because it makes it possible to witness quantum gravity effects at the laboratory scale. The approach was independently proposed by S. Bose and collaborators, and by C. Marletto and V. Vedral, \cite{MAVE17a, BOMAMO, MAVE17b}; the two teams had different rationales, but converged on the same proposed experimental scheme. 

The key idea is to test gravity with two quantum systems (probes). The original proposal has two masses each in a superposition of locations (but one could use other degrees of freedom, as we shall discuss in section 5). If gravity can entangle the quantum probes (under assumptions that we shall discuss later) then gravity must have some non-classical features -- meaning, informally, that it cannot be described by a fully classical theory such as general relativity. Hereinafter, we shall call this effect {\sl gravitationally induced entanglement} (GIE), \cite{MAVE17b}. In the literature, other names are found, such as the Bose-Marletto-Vedral effect, \cite{CHRO19} or Gravitationally Mediated Entanglement. The new and surprising element that the GIE proposal brings about is that signatures of quantum gravity can be detected below Planck's scale. Indeed, the GIE experiment has the tantalising feature of using masses of a few orders of magnitude {\sl smaller} than Planck's mass -- of the order of nanograms. This aspect makes it closer to experimental realisation than any previously proposed experiment. 
The reason why this is possible is that the experiment probes gravity's quantum side {\sl indirectly}, using the two quantum probes, as opposed to measuring directly quanta of the gravitational field predicted by the quantum theories of gravity, such as gravitons. 

Throughout, instead of `quantum effects', it is more accurate to talk about {\sl non-classicality} of gravity, following the information-theoretic definition introduced in \cite{MAVE20}. A system is `non-classical' if it must have at least two distinct physical variables that are necessary to describe its features, and yet cannot be measured to arbitrarily high accuracy simultaneously (i.e. by the same measuring system). These variables in quantum theory correspond to incompatible observables, as prescribed by Heisenberg's uncertainty principle, but the notion of non-classicality is more general than quantum theory. In section 4 we discuss how non-classicality can be defined exactly within the constructor theory of information, \cite{DEUMA}, which generalises quantum information theory to the domain where quantum theory may not apply.

{Here it is important to open a digression to explain what one means by a ``theory of gravity". The current best explanation of the gravitational interaction, provided by general relativity, considers gravity as a {\sl mediator}, with independent degrees of freedom, that propagate signals at a finite speed (the speed of light). We call gravity a ``mediator" and not a field here because there is an open discussion about whether it is appropriate to consider gravity a field; however, this discussion is not relevant here. Theories where gravity is a mediator can then describe it as classical, or as non-classical. In the former category, in addition to general relativity, we find semiclassical gravity, \cite{ROS, MOL}, and in particular quantum field theory in curved spacetime, \cite{BIRDAV}); in the latter, we find the many theories of quantum gravity that have so far been proposed (see \cite{KIE}). In contrast, Newtonian gravity is a theory where the gravitational interaction happens via instantaneous action at a distance: it is a classical theory of gravitation, where the gravitational interaction is not mediated. One can also retrieve Newtonian gravity as a limit of general relativity, and it can serve as a useful tool to perform computations in certain domains of applicability, \cite{KIE, Donoghue, WALLACE, CAR}. It is important to stress that the detection of GIE is a method to provide an experimental witness of the fact that gravity, intended as a mediator with independent degrees of freedom, is non-classical. We shall therefore call the protocol to detect GIE a {\sl witness of non-classicality}, \cite{MAVE17b}. The word `witness' originates in the context of quantum information theory, designating a Hermitean operator that can be measured on a quantum system to test experimentally that it is entangled to another system, \cite{HORHOR}. In general, successfully measuring a witness is a sufficient {\sl but not necessary} condition for two systems to be entangled: there are entangled systems that do not satisfy the witness criterion. Here we use the term in a different context, but with a similar meaning: successfully carrying out the protocol to witness non-classicality, thereby measuring GIE, is sufficient evidence that gravity is non-classical. However, failing to detect GIE does not immediately imply that gravity is classical, as we shall discuss in section 4.}

There are two notable features in the GIE non-classicality witness. 

One is that it probes a regime where quantum-gravity effects are relevant and measurable, but general relativity's effects are not, because gravity can be described in the Newtonian limit of general relativity. {GIE has now revealed that quantum effects in this regime are far easier to access experimentally, a fact that had so far gone unnoticed.} 
 
The other is that the GIE witness is based on a general theorem (the `General Witness Theorem', (GWT), see section 4) that can be proven with minimal assumptions about gravity's dynamics, \cite{MAVE20}.  This, in short, is the theorem: if a system  ${\bf M}$ (e.g. gravity) can entangle two quantum systems ${\bf Q_A}$ and ${\bf Q_B}$ (e.g. two masses) by local interactions (by ${\bf Q_A}$ coupling to ${\bf M}$ and ${\bf M}$ to ${\bf Q_B}$, but without ${\bf Q_A}$ and ${\bf Q_B}$ interacting directly) then ${\bf M}$ must be non-classical. The theorem, in this general form, is a generalisation of the Local-Operation-and-Classical-Communication (LOCC) theorems of quantum information, \cite{HORHOR} (which assume quantum theory's formalism) to the domain where quantum theory may not apply. Other versions of this theorem, based on assuming more restrictive assumptions, e.g. the formalism of quantum theory, appeared in closely related works, \cite{PAT17, MAVE17b, BOMAMO}.

In short, the GWT rests on two principles. One is the principle of locality, or {\sl no action at a distance}, that has to be satisfied by the dynamical theories describing both the entanglement mediator and the probes. (This principle then allows one to assume pairwise interactions of each quantum probe with the mediator, but not directly with each other).  
The other is the principle of {\sl interoperability of information}, \cite{DEUMA}. Informally, it means that the composite system of two systems that can each contain information must be capable of containing information, too. For instance, two systems each capable of containing one bit of information, when considered jointly can hold two bits worth of information. Applied to GIE, this requirement imposes that, irrespective of the particular theory of quantum gravity adopted to model GIE, particular interactions must exist between the two quantum probes and gravity (even when the quantum probes are each initially in a superposition).

Due to its generality, the GWT is akin in spirit to Bell's theorem, \cite{BELL}, which has been a cornerstone of ruling out classical models for quantum statistics: by violating Bell's (and related) inequalities in an experiment, one can rule out a vast class of local hidden-variables, real-valued stochastic models; likewise, when observing GIE, one can rule out all classical theories of gravity obeying the above-mentioned general principles, \cite{MAVE18}: classical theories that are known (e.g., quantum field theory in curved spacetime, general relativity, collapse models -- see section 4) and those that are yet to be discovered. 

The generality of the information-theoretic approach has a cost: it is impossible to use the GIE proposal in order to refute particular theories of quantum gravity against others. (Of course, one could modify the experiment so as to look at higher order corrections to the phases generating the entanglement, but these appear to be more difficult to access experimentally). In the quasi-Newtonian regime where GIE is generated, existing non-perturbative proposals such as string theory and loop quantum gravity agree with the predictions of linearised quantum gravity, \cite{KIE}, hence it is impossible to claim that this experiment, should GIE be observed, discriminates any particular quantum gravity theory from another. 
 
Indeed there is a lively debate about what observing GIE would imply exactly, which we shall discuss in section 4. For instance, some authors claim that it would not after all imply the non-classicality of gravity, because it only probes gauge degrees of freedom of the gravitational field, \cite{ANHU}; or because there are hybrid quantum-classical models that can generate GIE nonetheless, \cite{HARE18}. Other authors have claimed instead that observing GIE would even confirm the existence of gravitons, \cite{MAMABO}. 

The GIE experiment is at present only a theoretical proposal, and significant technological challenges remain in realising practical implementations of the protocol. In section 5, we shall discuss the main variants of the experiment that have been proposed in order to make it simpler; while in section 6 we discuss the main ongoing experimental efforts and their challenges. Despite the challenges, the GIE proposal however is extremely promising and it would constitute, if GIE were to be observed, the first experimental refutation of Einstein's theory of general relativity {(as a classical theory of gravity)}. It would imply that the latter needs to be corrected to take into account quantum features, as proposed in existing quantum gravity theories. If GIE were not to be observed, a number of other interesting possibilities would also open up. It may be that gravity is classical after all, and that quantum theory is false beyond certain scales. It may also mean that gravity is quantum, but the theories we are using to predict the effect are inadequate. It may also mean that the principles underlying the witness could be false. Any of these possibilities would be of great interest for the advancement of fundamental theoretical physics. 

In this review, after discussing some historical background (section 2), we shall proceed with describing the scheme of the proposed experiment for detecting GIE (section 3); its theoretical underpinnings and the implications of observing GIE (section 4), its proposed variants (section 5). In section 6 we shall review the contemporary experimental efforts and the main challenges for the GIE experiment to be implemented, and in section 7 we shall given an overview of a few exciting open problems in this nascent field.

\section{Brief historical background}

The GIE experiment is the latest in a long tradition of thought and actual experiments aimed at testing quantum aspects of gravity. For an extensive review of the philosophical and historical aspects of such experiments, see \cite{HUG}. {Here we only summarise briefly the major conceptual milestones} in the search for experimental confirmations of quantum gravity. 

\subsection{Milestones towards testing quantum gravity}

Historically, the conference that took place in Chapel Hill in $1957$, on quantum gravity, was particularly fruitful and set the tone for quantum gravity debate in many years to come. In particular, there was an important discussion between R. Feynman and H. Bondi, \cite{CHAP}. The key issue was how one would discriminate between genuine quantum superpositions and mere statistical mixtures of different gravitational states. In reply to Bondi's comments about the difficulty of testing quantum gravity, Feynman proposed a thought experiment involving a single mass that could undergo interference (figure 1), and pointed out that if gravity is quantum, then it should get entangled with the mass during the interference experiment, \cite{MAVE17a}. Feynman designed this experiment in order to refute statements by Bondi that even a mass prepared by classically randomising its position could somehow elicit the same response from the gravitational field as a quantum-superposed mass, thereby making quantum gravity potentially irrelevant. Feynman contradicted Bondi, stating that interference is a characteristically quantum phenomenon, which cannot be emulated by stochastic theories. If one could have the gravitational field itself participate in an interference experiment, Feynman argued, it would be beyond doubt that it must be quantum. Performing interference with the gravitational field would be equivalent to detecting interference of a single photon for the electromagnetic field. 
{Note that this thought experiment is different from the calculation included by Feynman in \cite{FEY3, Feynman}, to compute the gravitational interaction between two masses based on the exchange of gravitons. There, Feynman does not talk about the entanglement arising due to this interaction, although of course that calculation can be used to model that effect; also, that two-masses scenario is not used to argue for the quantisation of gravity. }

The single-mass experiment, as suggested by Feynman, would not allow one to conclude that gravity is quantum, even if performed successfully. For the single-mass experiment would be compatible with a classical gravitational field coupling with the quantum mass -- it is the mass, not the gravitational field, that undergoes interference in that proposed experiment. The GIE experiment, as we shall explain, goes one step beyond the experiment proposed by Feynman, using two masses each in a superposition. This is what makes the GIE experiment capable of capturing gravity's non-classicality.  

\begin{figure}[h]
	\centering
	\includegraphics[scale=0.4]{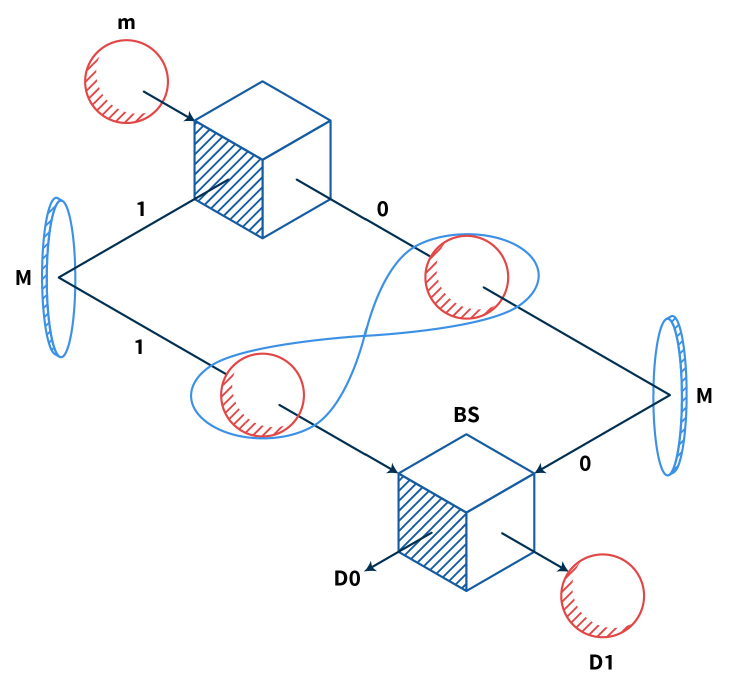} 
	\caption{{Adapted from Feynman's thought experiment, \cite{CHAP}, where a single mass undergoes interference, here represented in a Mach-Zehnder interferometer. This situation is in contrast to the state of a classically mixed mass, where interference is not permitted.}}	
\end{figure}

The next relevant experiment was proposed, and performed, by D. Page and C. D. Geilker, \cite{PAGGEI}. {These authors crystallised in an actual experiment a number of observations and arguments, intended to rule out a particular approach to semi-classical gravity,} {which improved on {\sl quantum field theory in curved spacetime}, \cite{BIRDAV}, by including the back-reaction of the gravitational field on a superposed mass. In particular, this version of semiclassical gravity treats the gravitational field by averaging the stress-energy tensor in the state of the quantum mass.}
The experiment consists of a quantum particle that controls the preparation of a Cavendish device, in one of two configurations: one has a counterclockwise torque (represented here by the state$\ket{c}$), the other configuration (represented here by the state$\ket{a}$), has the opposite torque (see figure 2).  Assuming the controlling particle is a qubit, and that its computational basis is used as control variable, the state of the joint system of the Cavendish device and the controlling particle is then:
\begin{equation}
\ket{\psi}= \frac{1}{2}\left(\ket{0}\ket{c} +\ket{1}\ket{a}\right)\;.
\end{equation}
In this state, there is a mass distribution where the two masses of the Cavendish device are in an entangled state with the qubit controlling the preparation. Quantum field theory in curved spacetime predicts that the gravitational field associated with this mass distribution is, in fact, an average of the two gravitational fields, each generated by one of the configurations. Hence, it predicts that the torque on the Cavendish device is identically zero. Whereas if one assumes that gravity is quantum, then there is a non-zero torque in each of the branches of the global quantum state including the quantum particle controlling the preparation and the Cavendish device itself. 

\begin{figure}[h]
	\centering
	\includegraphics[scale=0.4]{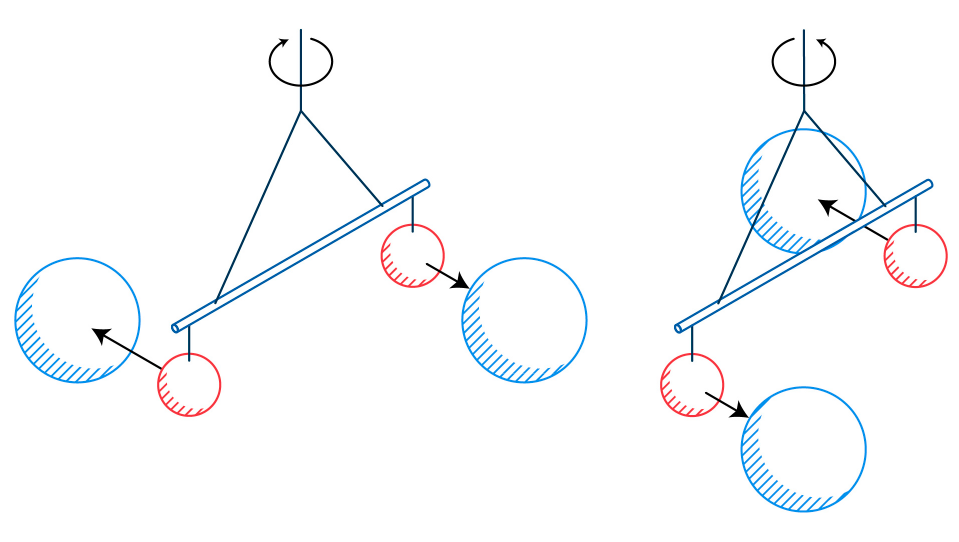} 
	\caption{The Page and Geilker experiment: two possible configurations of the Cavendish device. On the left, the clockwise configuration $\ket{c}$; on the right, the anticlockwise configuration $\ket{a}$.}
\end{figure}

The experiment was performed by using a radioactive decay process, which had a $1/2$ probability to decay after a 30 second interval, and $1/2$ probability to decay after a longer interval. In the former case, the device was prepared in one configuration (e.g. clockwise); in the latter, in the opposite configuration (e.g. anti-clockwise). Upon performing the experiment, an actual torque was measured in each round of preparation, and the authors conclude that this rules out quantum field theory in curved spacetime. However as the authors themselves recognise, the evidence provided by the experiment is merely consistent with the fact that gravity is quantum, but does not rule out other classical descriptions of gravity. Indeed, any gravitational collapse theory would also be compatible with what is observed in the experiment. So would the preparation of the Cavendish device with a classical coin being tossed, with no quantum effects being involved on the controlling particle's side: this would be a fully classical model for gravity and the masses involved. This experiment simply has nothing quantum in it: the controlling particle has already undergone decoherence by the time it controls the device. The experiment as performed shows however that quantum field theory in curved spacetime, if interpreted {\sl ad litteram}, cannot be a fundamental description of physical reality, as it leads to unphysical conclusions in certain situations, and it can be therefore easily ruled out. Despite this, quantum field theory in curved spacetime is still useful to perform approximate calculations in particular regimes, such as in the case of the Hawking radiation, \cite{BIRDAV}.

Another important milestone was the experiment performed by R. Colella, A. W. Overhauser and S. A. Werner (COW), \cite{COW}. This experiment is a neutron interferometer whose fringes are affected by the Earth's gravitational field, see figure 3. {Specifically, a neutron is set into a superposition of two locations. Let one of the arms of the interferometer be indicated by $0$ and the other by $1$.  After the first beam-splitter the neutron in the state $\left(\alpha\ket{0}+\beta\exp(i\phi_g)\ket{1}\right )$, {where $\phi_g=\frac{2\pi g m_n}{h^2}\lambda A\sin(\phi)$} is a relative phase that depends on the tilt angle $\phi$ of the plane of the interferometer with respect to the horizontal position (parallel to the Earth's surface), on the mass $m_n$ of the neutron, and we have introduced $g$ the gravitational coupling constant, $h$ Plank's constant, $\lambda$ the initial De Broglie wavelength of the neutron,  $A$ the area enclosed by the interferometer. The relative phase has been measured for different values of the tilt angle and to different degrees of accuracy.}

\begin{figure}[h]
	\centering
	\includegraphics[scale=0.4]{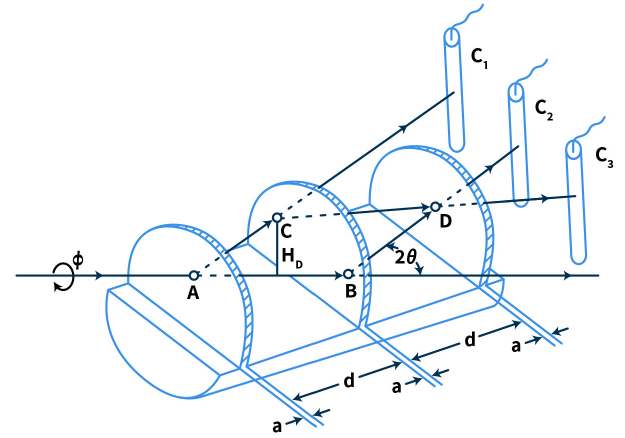} 
	\caption{{The COW experimental setup. A neutron undergoes diffraction, modulated by the tilt angle with respect to the Earth's surface. The phase shift induced by gravity has been measured for various values of the tilt angle.}}
\end{figure}
The importance of this experiment is that it shows that a mass superposed across different locations does couple with gravity at the scale of the experiment, respecting linearity and it does not undergo collapse. {In other words, a classical gravitational field coupled with a quantum source behaves like any other external field in non-relativistic quantum mechanics.} Hence, at that scale, collapse-based models due to gravity are ruled out, because the neutron preserves its coherence. However, this experiment could be explained by a fully classical model of the gravitational field, as it is customary to do in the theoretical analysis of the COW phase. Hence it does not provide evidence for quantum gravity for much the same reason as Feynman's proposal. 

There are also several proposed and performed experiments testing the wave-function collapse due to gravity. A good review of these is in \cite{COLLAPSE}. The difficulty with these experiments is that even a fully unitary model of decoherence would in fact explain the observed phenomena (see section 4.3), so they cannot be taken as proving that gravity causes any collapse at the relevant scales. In other words, even the full quantum gravitational field could be compatible with observing decoherence of massive superpositions.

\subsection{Arguments pro or contra quantum gravity}

There are several arguments that suggest that gravity's quantisation is untenable and collapse of the masses' wavefunction must occur at a certain scale. Such arguments are based on a number of different rationales. One of the most prominent arguments is that quantum physics, when applied to gravity, might violate the equivalence principle in one of its existing forms. {These issues emerge when a quantum source, such as a mass in a superposition of different locations, is coupled to a classical background.} For instance, the weak equivalence principle states that all objects, independently of the mass, accelerate the same way in the gravitational field. However, in quantum mechanics, the de-Broglie wavelength of a particle depends on its mass. Therefore different masses, when dropped in say the Earth's gravitational field, will produce different fringes. This seems to be a violation of the equivalence principle, \cite{DAV}. { In \cite{VIOLA} the consequences of the equivalence principle are studied for a test particle in free fall, prepared in quantum states with or without classical analogues. Another version of the quantum equivalence principle is also suggested in \cite{ANAN} and its application to a particular cosmological models with topological defects called 'cosmic strings' is proposed. Finally in \cite{GREEN} it is emphasised how the equivalence principle problem epitomises the fundamental disconnect between quantum theory and theories of gravity.}

However, one must be careful when generalising classical principles (such as say the energy or momentum conservation, or indeed the equivalence principle) to quantum mechanics. {There are in fact many different versions of proposed quantum equivalence principles, \cite{HAR, ZYCBRU, ANHU1}}. One could argue that, like other principles originally formulated within a classical theory, Einstein's equivalence principle should be extended to quantum systems by considering their classical version as applicable in each branch of a superposition, \cite{MAVE20a}. {Namely, if a massive particle is spatially superposed, it generates two different gravitational fields at a given point (assuming that the gravitational field is treated quantum-mechanically and in the first linear order of approximation, so that the same standards apply as in quantum electrodynamics). A test particle located at that distant point would then accelerate in both branches, towards the massive particle's respective locations. The state of the initial mass, the field and the test mass would then be:
\begin{equation}
\frac{1}{\sqrt{2}}\left (|r_1\rangle |g_1 \rangle |a_1 \rangle +|r_2\rangle |g_2\rangle |a_2 \rangle\right)
\end{equation}
where $g_i$ is the state of the field corresponding to the massive particle being at position $r_i$, while $a_i$ is the corresponding acceleration of the test particle. 
The equivalence principle, which says that the gravitational field is indistinguishable (locally) from acceleration, applies in each of the superposed spatial branches. {If the branches are not represented by orthogonal states, but are non-orthogonal states, it is still possible to rewrite the overall state as a sum of orthogonal branches, and then apply the principle to each of them. This is no different from what happens in the case of other principles such as the conservation of energy.} This is to be expected since each branch represents a classical gravitational scenario, where the position observable is sharp with the respective values. It is of course possible that this view of the equivalence principle will be experimentally invalidated (we do not have any experimental evidence in this domain to guide us); however, there is no prima facie reason to think that the equivalence principle is in conflict with quantum physics (any more than energy conservation is). }

In contrast to those arguments against quantum gravity, there is also a long-standing tradition of arguments to conclude that gravity must be quantised, purely on theoretical grounds. Two important cases stand out in the historical literature: one is DeWitt's argument for what DeWitt called the `totalitarian' property of quantum theory \footnote{D. Deutsch, private communication.}, to conclude that if a quantum system can interact with another system, then the latter must inherit the quantum uncertainty of the first system too \cite{DEW, DeW-Book}. This argument{(which is an extension of the original argument by Bohr and Rosenfeld about the EM field, \cite{BU})} is phrased assuming the full machinery of Lagrangian mechanics, and thus has been criticised for being too restrictive. Marletto and Vedral have generalised it to a framework (constructor theory) where only general principles need to be assumed instead of unitarity, \cite{MAVE17c, MAVE17d}. These arguments are important here because they are the precursors of the GWT - the general witness theorem supporting the GIE protocols to test quantum gravity.

The other important theoretical argument along these lines was proposed by K. Eppley and E. Hannah, \cite{EPHA}: they argue that if a quantum mass interacts with a gravitational wave, either there is collapse or the wave must also have some quantum features. This argument has been criticised, like DeWitt's, for being too narrow in its assumptions, and also for being inconclusive as it shows that some statistical uncertainty must spread to the classical side, but without showing that it is quantum uncertainty, \cite{KIEREG, MATT}. 

{Finally, there has been a long history of studies on the consistency requirements for hybrid quantum-classical models, \cite{ TERNO, CONST}. In these arguments it is shown that hybrid classical-quantum models can exist, but at the cost of breaking some fundamental principles such as conservation laws or locality. As we shall discuss later, this logic is similar to the argument supporting the idea that the GIE detection as a witness of quantisation of gravity. The current GWT is a generalisation of these arguments where the formalism is not assumed to be phrased within a particular dynamical model such as Hamiltonian or Lagrangian mechanics.}

\subsection{Arguments for the impossibility of detecting gravitons} 

In this context it is important to mention the arguments demonstrating the impossibility of detecting quanta of the gravitational field, also called `gravitons'. Gravitons are a prediction of ``Linear Quantum Gravity'' (LQG), which is a canonical scheme to quantise gravity following a linearisation procedure of Einstein's equations, \cite{KIE}. 
{Here by `linear' we refer to the linear approximation of the metric field that is deployed in order to solve Einstein equations approximately. In this approach, the metric is approximated by small perturbations of the flat spacetime metric, and then substituted into Einstein's equation, which are then approximated to the first order in the perturbation of the metric. This leads to an approximately linear set of equations, whence the name for the approach. }

Indeed, it was long hoped that one could apply the same quantisation procedures to gravity as those working for electromagnetism, and that the same way one can detect single photons in quantum electrodynamics, one could detect single gravitons in quantum gravidynamics (or geometrodynamics as Wheeler called it, \cite{WHE}). Neither hope was fulfilled: the standard techniques that apply to the electromagnetic field fail with gravity unless one proceeds with linearising Einstein's equation {and then breaks explicit Lorentz-covariance with an extra approximation step, \cite{CAR}}, thus landing on LQG, \cite{KIE}. Moreover, the predictions of LQG are discouraging as far as detecting gravitons, \cite{DYS}, as we shall now briefly recall. {Still, there are models that suggest the possible amplification of quantum gravitational effects beyond the Planck scale. For instance, a different setting in which quantum gravity effects may operate in macroscopic scale may be near-horizon regions of black hole, \cite{ASHT}.}

All arguments for the impossibility of detecting gravitons, {such as \cite{DYS}}, rest on the fact that the gravitational fine structure constant $\alpha_G = G m_e^2/\hbar c = (m_e/m_P)^2\approx 10^{-45}$ (where $m_e$ is the electron's mass and $m_P$ is Planck's mass), which determines the strength of coupling of gravity to matter, is much smaller than the fine structure constant $\alpha \approx 1/137$, which determines the strenght of light-matter interactions. This means that if an excited atom takes a nano-second to emit a photon, it would take $10^{36}$ seconds to emit a graviton. This is nineteen orders of magnitude longer than the age of the universe! 

Two other arguments are worth quoting here. One is due to G. Baym and T. Ozawa who discuss the possibility of obtaining the which path information in a double slit experiment by detecting a graviton emitted in the process, \cite{BAOZ}. The gravitational energy difference between the mass $m$ being located at one of the slits or other (separated by $d$) and a distance $R$ away is $Gm/R^2 d$ and this needs to be at least as big as the energy of one graviton $\hbar c/d$ of wavelength no greater than $d$ (otherwise we would not be able to obtain the which-slit information). This gives us the following condition $(m/m_P)^2 \geq (R/d)^2$. However, the size of the interference fringes is then $l=L\lambda/d =\hbar L/mvd\leq \hbar LR/mvd\leq l_P$, i.e. the interference fringes separation is less than the Planck length. This means that in order to have the which path information due to the emission of gravitons, the fringes - in the case when no which-path information has been obtained - are too small to ever be observable (possibly being even beyond what is physically possible).

{A. Peres and N. Rosen, \cite{PERROS}, on the other hand, make a more general argument.  Their conclusion stems from a simple observation which can be stated as follows: there is a relationship between the mass $m$ and size $s$ of an object and the time $t$ it takes to produce the interference fringes through diffraction, $\hbar t \approx m s^2$ (this is derived by using the Fraunhoffer condition for interference $D\lambda \approx s^2$, and expressing the distance to the screen as $D=vt$ where $mv=\hbar/\lambda$). This relationship suggests that an object of the Planck mass and size of $10^{-4}$m would take longer than the age of the universe to interfere. No gravity is involved in this argument and the limit would apply in any spacetime geometry. Interestingly, the formula is satisfied exactly by the object of the mass of the universe, the size of the Planck length and interfering for the duration of the existence of the universe.} 
{There are also some arguments that indicate the possibility of detecting gravitons by other means, such as via decoherence, \cite{RIE}. However, in addition to still being experimentally challenging, such arguments assume to apply unitary quantum theory to gravity in order to conclude that decoherence occurs via gravitons, which is the point in question.
We finally note that the GIE experiment is a method to detect non-classical features of gravity without detecting gravitons directly (as we shall discuss later); hence these arguments are not contradicted by the possibility of detecting GIE. }

\section{The GIE experiment}

The GIE experiment utilises an interference experiment with two quantum probes (as opposed to one probe, like in Feynman's and the COW experiments) to perform a test of the non-classicality of gravity. In section 4 we shall define the notion of non-classicality exactly and without relying on the formalism of quantum theory, but for present purposes it is sufficient to know that a system is non-classical if it has at least two physical variables that are both necessary to described the system, and they obey Heisenberg's uncertainty principle -- i.e., in quantum theory, that do not commute. {The motivation for not relying on the formalism of quantum theory is that quantum gravity may not be expressed in that formalism, just like quantum mechanics and general relativity are not expressed in the formalism of Hamiltonians mechanics. Furthermore, if one wishes to convince the sceptic of gravity quantisation with an experiment, it is desirable to base any argument to do so on assumptions that do not involve applying quantum theory and its formalism to gravity itself. }

\subsection{GIE experimental scheme}

In the original proposal, two massive particles are prepared in a superposition state and let become entangled locally {\sl only} via gravity. By designing the initial states and controlling the masses appropriately, it is possible to produce a detectable entanglement with relatively small masses (at the nanogram scale). Such entanglement, when analysed in the presence of specific theoretical assumptions, implies that classical models for gravity (including general relativity) must be ruled out. Let us see in detail how this works. 

{\bf The experimental scheme.} Two quantum systems ${\bf Q_1}$ and ${\bf Q_2}$ with equal mass $m$ are entangled only via the gravitational field -- which plays the role of a mediator that we shall call ${\bf M}$. 
We can consider a simplified scheme with two interferometers, \cite{MAVE17b}, each located so that both masses are subject to the same Earth's gravitational field. In section 6 we shall explain that there are a number of possible different implementations of the interferometer. Bose and collaborators in \cite{MAMABO} discuss a possible setup with spin degrees of freedom being used to generate the superpositions of masses, however there are other possibilities -- the spin is in fact simply a label for the masses' position degrees of freedom, which are those getting entangled. Alternatively, one could have matter-wave interferometers, parallel to the Earth's surface, \cite{MAVE17b}; nanomechanical oscillators, or trapped particles \cite{KRITHAPAT}.

We shall consider a generic set up with an interferometer {with two different paths (also called `arms')}, each labelled by $0$ and $1$. Each mass is put in the state $\frac{1}{\sqrt 2} \left(\ket{0}+\ket{1}\right )$ by the first beam-splitter in the interferometer (see the figure 4).
\begin{figure}[h]
	\centering
	\includegraphics[scale=0.4]{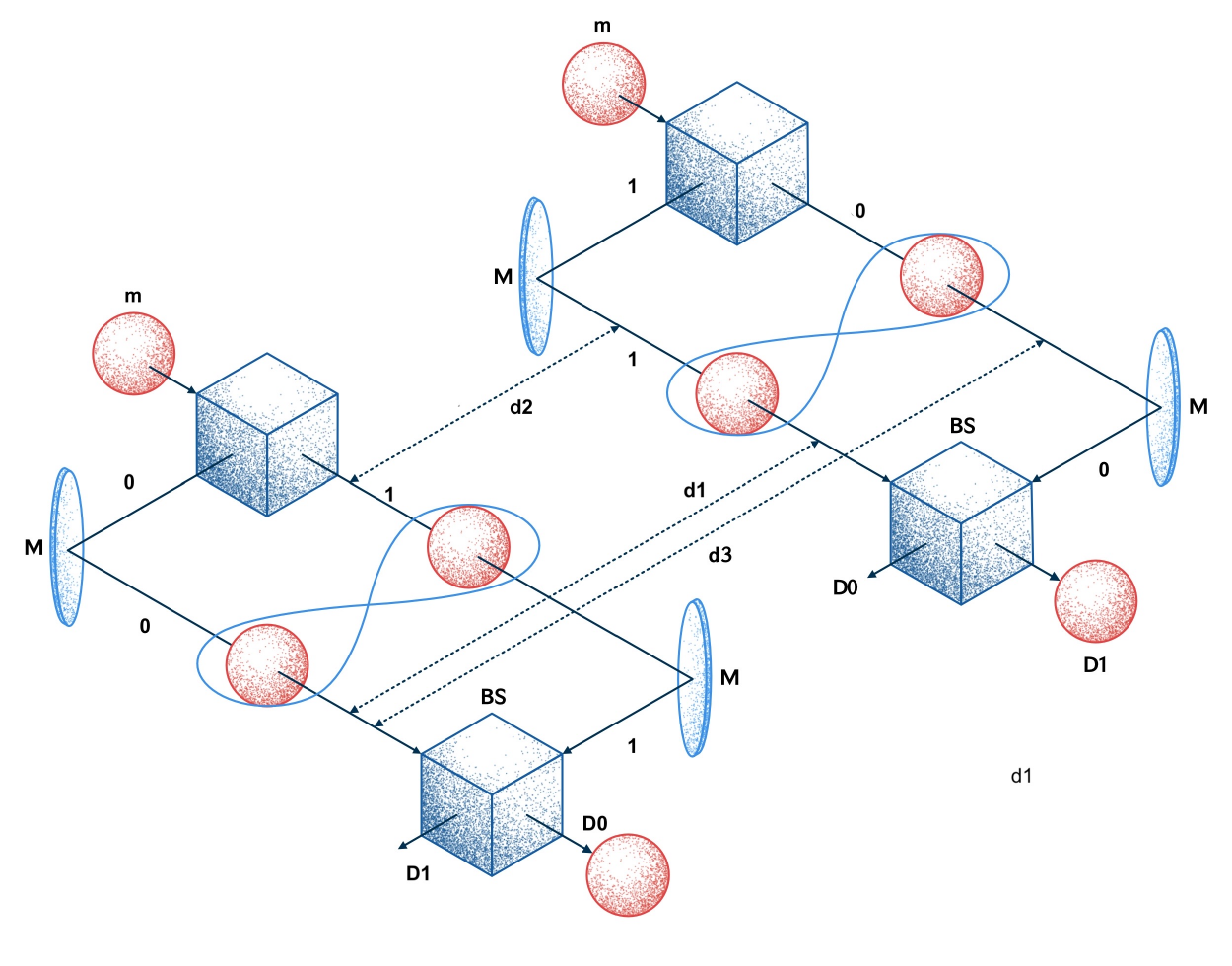} 
	\caption{{GIE generation with two masses. Each mass $m$ individually undergoes interference, and interacts with the other mass via gravity. BS is a beam splitter; M is a mirror; Di with $i=0,1$ represents the detector on path $i$. $L$ is the length of the lower arms of each interferometer. The distance between the lower arms of the two interferometers is $d_2$, the distance between the upper arm of one interferometer and the lower arm of the other interferometer is $d_1$. For simplicity, in the text we assume $d3$ to be much larger than $d1$ and $d2$, so the associated phase can be ignored.}}
\end{figure}

{The masses in different branches interact each via a different gravitational field configuration, which leads to phase differences between the four branches. Assuming for simplicity that $d_3$ is much larger than $d_1$ and $d_2$, and thus leads to a negligible phase, the state of the composite system becomes,} before they enter their respective final beam-splitters: 
\begin{eqnarray}
\frac{1}{2}\ket{0}\left(\ket{0}+\exp{(i\phi_1)}\ket{1}\right )+ \nonumber \\ 
\frac{1}{2}\exp({i\phi_1})\ket{1}\left(\ket{0}+\exp{(i(\Delta{\phi}))}\ket{1}\right )\;.
\end{eqnarray}

\noindent where $\phi_1$ and $\phi_2$ are the relative phases acquired by the masses due to the gravitational potential generated when they are, respectively, at distance $d_1$ and $d_2$ from one another; $\Delta\phi=\phi_2-\phi_1$ is their difference. We can also suppose for simplicity that the gravitational interaction of the masses on the two most distant arms is negligible. Assuming that the dominant contribution to the gravitational interaction is quasi-Newtonian, and that the general-relativistic contributions are negligible, the value of the phase is $\phi_i=\frac{m^2G}{\hbar d_i}\Delta t$; where $G$ is the gravitational coupling constant; $\Delta t=\frac{L}{v}$ is the time spent by each mass on the horizontal arm of the interferometer, of length $L$; and $v$ is their velocity. {Note that the phase is here calculated using the usual Newtonian potential, but as we explain in the next subsection this phase should be calculated with a LQG model that models the gravitational field dynamically, as a third, mediating quantum system.}

In each interferometer, the probabilities $p_{\alpha}$ for the mass to emerge on path $\alpha=0,1$ are:

\begin{equation}
p_0=\frac{1}{2}\left(\cos^2{\frac{\phi_1}{2}}+\cos^2{\frac{\Delta \phi}{2}}\right)\;,\\
p_1=\frac{1}{2}\left(\sin^2{\frac{\phi_1}{2}}+\sin^2{\frac{\Delta \phi}{2}}\right)\;.
\end{equation}

{The two masses are maximally entangled by the action of the gravitational field when $p_0=p_1=\frac{1}{2}$. This happens when $\phi_1=2n\pi$, $\Delta\phi={\pi}$ for some integer $n$. The state in this case can be written as $\frac{1}{\sqrt{2}}(\ket{0}\ket{+}+\ket{1}\ket{-})$ where $\ket{\pm}=\frac{1}{\sqrt{2}}(\ket{0}\pm\ket{1})$. Therefore to confirm entanglement, one would require to measure two complementary observables on each interferometer. If we measure the path of the first particle (the effective Pauli Z measurement), the second mass interferes with either the plus or the minus phase (i.e., it is in an eigenstate of the Pauli X). If, on the other hand, the second mass is first measured in X, there is no interference of the first mass, meaning that $X_1$ and $X_2$ are not correlated. In terms of probabilities: $p(0,+)=p(+,0)=p(1,-)=p(-,1)=1/2$ and $p(0,-)=p(-0)=p(1,+)=p(+,1)=0$. Therefore measuring the observable $X_1Z_2 + Z_1X_2$ will suffice to witness entanglement.} 

When instead $\phi_1=\Delta\phi=2n\pi$, the two masses are not entangled and each undergoes, separately, an ordinary interference experiment, emerging on path $0$ of the interferometer. For a fixed mass, by varying the arms' distance or their length, it is in principle possible to interpolate between those two cases, thus demonstrating all degrees of entanglement, ranging from no entanglement to maximum. Using the arguments discussed in section 4, this entanglement can be used as a witness that the gravitational field mediating the interaction is non-classical. In short, this is because if one were to use a classical mediator to couple the two quantum probes locally, the most general state that one can achieve by local interactions would be separable at all times, \cite{MAVE17b, BOMAMO, MAVE20}.

Feasibility considerations suggest that the experiment {is not too far from the reach of existing technologies}. For example, two superposed masses $10^{-12}$ kg interacting for a $\Delta t=10^{-6}$ s would achieve maximum entanglement, over distances $d\approx 10^{-6}$m. We shall discuss the main proposals to realise this scheme and their challenges in section 6. {We note that these are just orders of magnitude; for a more specific discussion based on various other feasibility considerations the reader can refer to section 6 and the references therein. Here we only mention that the main challenge is presented by decoherence, which has been thoroughly analysed in a vast literature -- see .e.g \cite{ZUREK, JOOS}.} 

Since any form of energy gravitates as far as general relativity is concerned, we could have other kinds of superpositions to probe the quantum nature of gravity. For instance, an object that is in a superposition of two different energy states would also produce two different kinds of the gravitational field (even though the object would occupy one and the same physical location). We shall discuss variants of the experiment in section 5.

\subsection{Linear quantum gravity model}

The phases that generate entanglement can be computed with LQG, \cite{MAVE17b, BOMAMO}, as we shall now recall.  

{We confine attention to the regime where Newtonian contributions are dominant, i.e. where the metric is represented by a small perturbation $\xi({\bf r},t)$ away from the Minkowski metric. In this regime, the spacetime metric can be approximated by $ds^2=-(1-2\xi({\bf r},t)/c^2)dt^2+d{\bf r}^2$, where $\xi({\bf r},t)$ is a field that may be time-dependent and propagating at speed no greater than $c$ (incidentally, all the conclusions of this review will also apply to the cases where gravitons have non-zero mass, as in e.g.\cite{DERHA}.)} 

(As we remarked in the introduction, this regime is different from Newtonian gravity, where the gravitational potential admits instantaneous action at a distance and it is static). In this regime the non-perturbative approaches to quantum gravity, such as loop quantum gravity and string theory, agree with the predictions of the linearised approach, \cite{KIE}; thus the proposed experiment would test their prediction, too, in this regime.  

We can model the gravitational field as a single quantum harmonic oscillator. In linear quantum gravity, $a$ and $a^{\dagger}$ can be interpreted as the bosonic annihilation and creation operators for gravitons. These degrees of freedom can be thought of as the scalar degrees of freedom that appear in the gauge of the Gupta-Bleuler formalism, see \cite{GUP, BOMAMO, MAVE23}. There are some mathematical subtleties to do with the metric in this formalism, but they are not relevant here for the computation of the phase.
The two masses can be initially modelled as two qubits -- whose z-component represents a discretised position of each mass - in this case, one of the two paths in an interferometer. Its eigenstates $\ket{a}$ where $a\in \{0,1\}$ represent the situation where the mass is on a definite path $a$; while $\ket{ab}$ describes the situation where the first mass is on path $a$ and the second on path $b$. 

We now compute the phases in the proposed experiment using the full linearised Hamiltonian. This is obtained from the general linearised Hamiltonian: $H^G_{int} =- \frac{1}{2} h_{\mu\nu} T^{\mu\nu}$, 
where $T^{\mu\nu}$ is the stress-energy tensor and $h_{\mu\nu}$ is the perturbation of the metric tensor $g_{\mu\nu}$ away from the flat (Minkowski) spacetime. The quantised gravitational field is then written in terms of the graviton creation and annihilation operators $a(k,\sigma), a^{\dagger}(k,\sigma)$, as:
\begin{equation}
h_{\mu\nu} \propto \sum_{\sigma} \int \frac{d^3k}{\sqrt{\omega_k}} \{ a(k,\sigma)\epsilon_{\mu\nu} (k,\sigma) e^{ik_\lambda x^\lambda} + h.c.\}
\end{equation}
where $\epsilon_{\mu\nu}$ is the polarisation tensor, $\sigma$ indicates two non-vanishing gravitational polarisations, while $\omega_k$ and $k$ represent the frequency and wavenumber of the relevant mode respectively (we are using the Einstein's convention of summation). 

In this experiment, the masses are non-relativistic and the stress-energy tensor simplifies to $T_{00} = m$. We can also consider, for simplicity, a single polarisation and a discrete sum over the relevant gravitational quantum modes.
The total Hamiltonian involving two masses and the gravitational field is therefore
\begin{eqnarray}
H & = & mc^2 (b^{\dagger}_1 b_1 +b^{\dagger}_2 b_2) + \sum_k \hbar \omega_k a^\dagger_{k}a_{k} \nonumber \\
& - &   \sum_{k,  n\in \{1,2\}} \hbar g_k  b^{\dagger}_n b_n (a_{k}e^{i k x_n}+a^\dagger_{k}e^{-i k x_n})
\end{eqnarray}
where the first two terms are the free Hamiltonians of the masses and the field respectively. 
We assume that the gravitation-matter coupling constant is given by
$g_k = mc \sqrt{\frac{2\pi G}{\hbar \omega_k V}}\;$, where $V$ is the relevant volume of quantisation (which will not feature in the relevant observables). The evolution of two masses of value $m$ at positions $x_1$ and $x_2$ interacting with the initial gravitational vacuum state can be solved exactly: $
e^{iH t} |m\rangle |m\rangle |0\rangle =\exp\{\hbar \sum_kV(k)t\} |m\rangle |m\rangle |\sum_k \frac{g}{\omega_k}(e^{-ikx_1}+e^{ikx_2})\rangle$, 
where $V(k)= \frac{g_k^2}{2\omega_k}(1+2\cos(-ik(x_2-x_1)))$.

When acting on the initially superposed state, this Hamiltonian generates entanglement between the two masses. The continuum version of the position-dependent part of $V(k)$, obtained by replacing the sum over $k$ by an integral, is:
\begin{equation}
{\rm Re} \left \{V \int dk \frac{4\pi Gm^2}{\hbar k^2 V}e^{-ik(x_1-x_2)}\right \} = \frac{Gm^2}{\hbar (x_2-x_1)}
\end{equation}
which provides the relevant quasi-Newtonian phase that can generate the relevant entanglement.  The key fact is that the above process relies on two complementary observables of the field, because the observables $\frac{1}{2i}(a-a^{\dagger})$ and $a^{\dagger}a$ are needed to generate the Hamiltonian dynamics, which is what makes the field non-classical and capable of mediating the entanglement. 

We can assume that the interaction between the masses and the field is `elastic', i.e., when the two masses are brought back to their original state, where each one of their positions is sharp, the field goes back to the original state, and it is unentangled with the masses. However, even if the interaction were not perfectly elastic, since the entanglement between the field and the masses is very small (at least for masses below Planck mass), the state of the two masses is approximately not entangled with the field at the end, thus leaving the field approximately unchanged. The same result can be obtained with the usual Lagrangian formulation of quantum field theory, where the interaction is established via the exchange of a single graviton between the two masses and the field, \cite{MAVE18, ROV2, MAMABO20b, BOMAMOSCH}.
The above Hamiltonian can be approximated via a quantum network model, \cite{MAVE18} which we briefly recall in appendix A. This gate model was simulated, together with decoherence, with NMR qubits in \cite{BHO}. A fully optical simulation of the same proposed quantum network model can be found in \cite{SCIA}. 

{\bf GIE gives physical meaning to Planck's mass.} The entanglement between the field and masses, which is necessary to generate GIE, can be quantified by the reduced entropy of the masses, and provides a novel insight into the physical significance of Planck's mass, as was suggested by Marletto and Vedral in \cite{MAVE18}. Since the field and the masses are weakly entangled, a good approximation of the reduced entropy is the linear entropy $S_L=1-{\rm Tr(\rho_{Q_i}^2)}$, where $\rho_{Q_i}$ is the reduced state of mass $Q_i$. {The magnitude of the reduced entropy is given by one minus the overlap between the two gravitational states squared as in: $1- |\langle\alpha_{ab}|\alpha\rangle|^2 = 1- \exp{(-\xi_{ab})} \approx \xi_{ab}\;$, where $\sqrt{\xi_{ab}}$ is a real-numbered shift depending on the exact coupling between the field and the masses, that causes the relevant phase-shift $\phi_{a,b}$ (see appendix B for more details).} This quantity could be very small compared to one, while still generate the desired entanglement between the two masses. Assuming the regime where the Newtonian contribution only is relevant, one has: \begin{equation}\phi_{ab}=w\alpha_{ab}=\frac{Gm^2}{\hbar d_{ab}}\Delta t=\left (\frac{m}{m_P}\right )^2\frac{c}{d_{ab}}\Delta t\;,\end{equation} where $\Delta t$ is the interaction time between the two masses, $m_{P}$ is Planck's mass, G is the gravitational constant, and $d_{ab}$ is the distance between the position $a$ of the first mass and position $b$ of the second mass. One can identify $\alpha_{ab}=\left (\frac{m}{m_P}\right )^2$. The entanglement between a spatially superposed mass and the gravitational field (if, indeed, it is quantum) would then offer another way of giving meaning to the Planck's mass. Namely, if we really want a spatially superposed mass to entangle appreciably to the surrounding gravitational field, according to the above formula, we need to engage masses on the order of and larger than the Planck mass. {In contrast, it is interesting that, as we discussed, maximal entanglement between the two quantum probes {\sl generated via the interaction with gravity} can be achieved even with masses that are smaller thank Planck's mass. This is the surprising effect that underlies the GIE phenomenon.}

\subsection{Gravito-magnetic model}

It is helpful to compare the LQG model with the analogous EM scenario. 

Following LQG, in the weak limit, the gravitational field of General Relativity resembles the electro-magnetic (EM) field. In this linear regime of weak gravity, the Christoffel's symbols of general relativity can be expressed in a form analogous to the electromagnetic field tensor, $F_{\mu\nu}$ - \cite{PERROS}. The electric-like components in the general relativity Christoffel tensor, $\Gamma_{\mu\nu}$, are the usual Newtonian gravity, while the magnetic-like components are those responsible for the so called inertial-induction gravitational effect (as we will see in more detail below). The latter is the gravitational version of Faraday induction and it has never been observed experimentally. The reason for its elusiveness is that the induced gravity by a moving mass is much weaker than the Newtonian gravity by the same mass.  

The degree of non commutativity of the electric-like and magnetic-like components of gravity can be 
understood by comparison with the EM case \cite{Bohr,Heisenberg,Heitler}. {The uncertainty relation for the electromagnetic field confined to a region of size $L$ is $\Delta E \Delta H\geq \hbar c/L^4$, where this holds for the components of the EM field in two orthogonal directions such as e.g. $E_x$ and $B_y$ (note that $H=B/\mu_0$)}. This degree of non-commutativity is already comparable to unity for confinements of the micron size $L\approx 1\mu$m (the Lamb shift, the Casimir effect and spontaneous emission are observable consequences of this non-commutativity). {There is a possible limitation on each individual gamma (unlike on the E and the B field components). This is due to the fact that increasing the mass of the probe will ultimately lead to a black hole, the equivalent of which does not exist in EM. This is not relevant to us because this latter limitation could be seen to be entirely classical. So what matters is the product of the uncertainties as in the EM field.}

For gravity, the uncertainty in the metric $g$ is of the order of $\delta g\geq l_P/L$, where $l_P$ is Planck's length \cite{Regge} (see also \cite{KIE}). Since $\Gamma \propto \partial g$ (where $\partial g$ is a generic derivative), the uncertainty in the components of the gravitational field is of the order of $\Delta \Gamma \geq l_P/L^2$. We therefore expect the electric-like and the magnetic-like components of $\Gamma$ to obey the following commutation relation in linear quantum gravity:
\begin{equation}
\Delta \Gamma_1 \Delta \Gamma_2  \geq \frac{l^2_P}{L^4} = \frac{\hbar G}{c^3 L^4} \; , \label{deltas}
\end{equation}
which is a relation already argued for by Peres and Rosen \cite{PERROS}. Note, however, that this is of the order of unity only if $L\approx \sqrt{l_P}$ which is smaller than the nuclear dimensions. Hence, we expect the effects of this non-commutativity in quantum gravity to be much weaker. In fact, some authors have claimed that none of the gravitational phenomena that mirror the EM case (such as spontaneous emission of gravitons, gravitational Lamb shift or gravitational Casimir effect) will ever be experimentally accessible.

However, one must emphasise that a) the analogy between gravity and the EM field exists only in the linear regime of GR; b) the crucial difference between gravity and the EM field is that in gravity anything that possesses energy acts gravitationally and is acted upon gravitationally. The former implies that in GR we also have to include the effects of gravitational energy on itself (``gravity gravitates") which makes the challenge of quantisation more formidable. The latter implies that there is a limit to how well individual components of the GR field can be measured (and not just their product, as for the electromagnetic field). Briefly, if we want to measure the components of the EM field more accurately, we just need to make our probe have more charge (and the back action on the field can be cancelled by a suitable choice of the interaction). However, the same trick fails in gravity since there is a limit to how massive something can be within a certain region and not become a black hole. The argument was originally used by Bronstein in 1936, \cite{Bronstein}, who was the first person to present a quantisation of the linear GR.

Consider {two like charges} that are moving parallel to each other. The force due to their electric fields will be repulsive, while magnetically, they will attract each other. The magnetic attraction is induced by the field of the moving charges (in a completely symmetric fashion in this case). The same is the case for gravity, other than the fact that the gravitational forces are always attractive. Imagine two masses $M$ moving parallel to one another at a distance $r$ as in our double interference proposal. 
Here, the electric-like gravity is given by
\begin{equation}
E_G = \frac{1}{\epsilon_G}\frac{M}{r^2} \; ,
\end{equation}  
where we use $\epsilon_G = 1/4\pi G$ in order to emphasise the electromagnetic analogy, while the magnetic-like component is
\begin{equation}
B_G = \mu_G \frac{I}{r} \; ,
\end{equation} 
where $I$ is the mass current density and $\mu_G = 4G/c^2$ (and, as in the EM case, $c^2 = 1/\epsilon_G\mu_G$, which is expected since the gravitational waves also propagate at the speed of light). These are the two components of gravity whose non-commutativity would be revealed by witnessing entanglement in our experimental proposal. 
Note that these two non-commuting components correspond to the two non-commuting degrees of freedom $a^{\dagger}a$ (representing the energy of the gravitational field per mode) and $a+a^{\dagger}$ (representing the quantised perturbation of the metric per mode) in the linear-gravity Hamiltonian presented in the earlier section; both are necessary for establishing the entanglement between the masses. The linear quantum gravity model, therefore, perfectly reflects the intuition based on the EM analogy with weak gravity. In quantum electrodynamics {it is also true} that the energy of the field does not commute with different field components. It should be said that the Christoffel-based commutation relations of the gravitational field can formally be derived in this linear regime and they would fully corroborate \eqref{deltas}, see e.g. \cite{PERROS}.
The following intuitive argument relying on consistency with ordinary quantum physics shows why these two components of gravity should be complementary. The resulting forces from the orthogonal $E$ and $B$ fields are co-linear (just like the Lorentz force between two charges moving in parallel). If we want to measure them simultaneously, we need a test charge which will respond to them. From the above formulae we obtain: 
$\Delta E_G \Delta H_G \geq \frac{G\hbar}{c^3r^3 L}$, which agrees with the above $\Delta \Gamma_1 \Delta \Gamma_2$ formula when $L=r$. 
However, as we said, the gravito-electric and gravito-magnetic components are weak in comparison with EM. In the GIE proposal the two masses should be {of the order of 1} nanogram, separated by a distance {of the order of 1 $\mu$} in order for entanglement to be experimentally observable. Suppose that the masses are both moving at speeds of about a million meters per second (we are assuming the highest speeds that are still non-relativistic to a good approximation). The force experienced by each mass $M$ is the gravitational Lorentz force given by $F=M(E_G + v \times B_G)$ and is the sum of the electric- and the magnetic-like components. For our experiment the electric-like force is of the order of  {$10^{-24}$ Newton} (one yoctoNewton), while the magnetic-like force is three orders of magnitude smaller. Again, it should be noted that, although each could be hard to detect in practice (although, see \cite{yocto}), their non-commutativity can in principle be detected indirectly by GIE.

The non-commuting degrees of freedom that are uncovered through the GIE would ultimately depend on the exact approach taken when modelling the interaction between the masses and the gravitational field. In the canonical approach, as we have just seen, it is the quantised metric tensor (which is the gravitational analogue of the electromagnetic vector potential) and the quantised components of the Christoffel symbol (which are the gravitational analogues of the $E$ and the $B$ fields). In the Hamiltonian/Lagrangian description of the interaction it is the energy of the field which, as an operator, does not commute with the operator that couples to matter (the metric tensor). In other approaches to gravity, such as say loop quantum gravity, it is the operators pertaining directly to the geometrical features that are no longer complementary. Hence, the operator for the volume does not commute with the operator for the area. However, these are not the operators that necessarily describe directly the interaction between the gravitational field and matter. In loop quantum gravity, the interaction could always be rewritten in terms of the volume and area operators. In string theory, where the fundamental entities are strings, there may well be another set of fundamental entities that give rise to the metric tensor and Christoffel symbols, but what exactly they are (presumably observables pertaining directly to strings) is not immediately relevant so long as they are non-commuting (otherwise they could not explain the gravitationally induced entanglement).

Just to illustrate how string theory would lead to the same conclusion as far as GIE, we present a sketch of the calculation here. The amplitudes $A$ are given by extending Feynman's sum-over-histories to a sum over string world-sheets interpolating between the string configurations, \cite{OHN}:
\begin{equation}
A \propto \int D(\textrm{worldsheets})e^{i \textrm{Area}} \; .
\end{equation}
The Area of the worldsheet swept out by the strings as they move through spacetime plays the role of the Action in standard quantum field theory. When this is applied to two input two output particles whose scattering is mediated by exchanging another particle this becomes the so-called Virasoro-Shapiro amplitude, whose poles indicate all the possible masses of mediating particles. One of these is the spin-2 $m=0$ particle, which is the graviton. There is then a general argument that goes back to Feynman: any theory of an interacting spin two massless particle must describe gravity. So string theory ought to reproduce gravitational physics at least at this level of approximation. {More generally, we expect that the high-energy descriptions such as string theory must reproduce in the low-energy limit the predictions of Einstein's gravity and quantum field theory. There are several arguments that support this idea, see e.g. for a pedagogical review \cite{Donoghue}.}

\section{Implications of the GIE experiment} 

A complex debate on what observing GIE implies has arisen since its proposal.  {One possible way of regarding the detection of GIE is that it confirms the predictions of existing quantum gravity models, which all concur in predicting the GIE phase in the linear and non-relativistic regime (albeit with considerable differences in the underlying mechanisms, \cite{CAR})}. However GIE does not show the validity of any of these theories. Indeed, it may well be the case that a quantum gravity model with no gravitons can also be corroborated by observing GIE experimentally. 
A more general way to express the implications of the experiment is that GIE rules out all classical theories of gravity that satisfy the general principles assumed by a general theorem about the witness of non-classicality on which GIE can be based. In this sense detecting GIE is agnostic about which quantum gravity sthould be used to model quantum effects in the gravitational field, leaving the possibility open that other systems than gravitons may be the right answer. 

The General Witness Argument allows the detection of GIE to be elevated to conclusive evidence that no classical model satisfying the assumptions of the theorem (which is a vast class of plausible physical models for gravity, including general relativity) is suitable to describe gravity.  In this sense detecting GIE is very similar to an experiment which detects a violation of Bell's (and related) inequalities, which allows one to rule out all hidden variable models for the experiment in question. According to this interpretation, detecting GIE is a final nail in the coffin of classical and semiclassical theories of gravity. In this section we proceed to discuss the different possible interpretations in some detail.

\subsection{Testing linear quantum gravity versus specific classical variants} 

One way to think of the GIE experiment is that it tests linear quantum gravity (see section 3) against a specific classical version of it, \cite{BOMASC, MAMABO}. If one adopts this viewpoint, the detection of GIE rules out general relativity, quantum field theory in curved spacetime, certain collapse-models that predict collapse at the relevant scale, and a decohered version of linear quantum gravity where gravity is an entanglement-breaking channel, \cite{KATAMI}, {as well as other stochastic semiclassical hybrids such as \cite{OPP}}. {If intended in this sense, the GIE experiment is a natural extension of works where one seeks to discriminate, within a specific quantum gravity model, the situation where gravity is quantum coherent, and where it is a decohered channel, see for instance \cite{KATAMI, ADD1, ADD2, ADD3}.}

{It is also important to notice that all such theories assume that gravity is a mediated interaction, and they are not action-at-a-distance theories. This is an extra assumption that must be either theoretically assumed or independently tested via other experiments.} 
Given that non-perturbative quantum gravity theories (such as loop quantum gravity and string theory) have linear quantum gravity as their low-energy limit, they too would be compatible with observing GIE. 
Some authors have argued that detecting GIE provides evidence for stronger claims. For instance, Bose and collaborators have argued that GIE provides evidence for the existence of gravitons as mediators of the entanglement formation process, \cite{BOMASC}. However this may not be the case. Linear quantum gravity is known to be an approximation scheme, that is problematic in that it starts from a linearised version of general relativity. Moreover, that linear quantum gravity is compatible with the detection of GIE does not mean that it is verified by observing GIE. It is simply one model that fits the data, but we know from independent reasons that it is not the ultimate quantum gravity theory. In addition, the GIE experiment does not test directly the existence of gravitons -- it is not the equivalent of observing a graviton being emitted or absorbed. {Indeed it is tempting to make parallels between the GIE experiment and the experiments that accompanied the discovery of quantum theory in the past century. In this context, one could think that a possible analogy is with the photoelectric effect. However, the photoelectric effect is based on detecting photons directly, whereas GIE only tests the presence of non-commuting degrees of freedom in gravity, not the presence of gravitons. Moreover, the photoelectric effect can also be modelled without photons, \cite{SCULLY}. So a better analogy could be with the detection of the blackbody radiation: the detection of blackbody radiation was an experimental refutation of the theory of classical electromagnetism. That effect however is still imperfect as an analogy; the important point is that it could not confirm the existence of photons, but it did show that the classical model was inadequate. Likewise, detecting GIE does not confirm the existence of gravitons, but it rules out all theories of classical gravity obeying certain assumptions. }

It has also been claimed by M. Christodoulou and C. Rovelli that the GIE experiment shows that superpositions of spacetime geometry must exist, \cite{CHRO19}. It is important to clarify that, strictly speaking \cite{MAVE20}, spacetime geometry during the experiment is never prepared in a superposition of its classical configurations. It is in a mixed state. {Indeed, consider a simple model with two qubits, one representing a superposed, and the other the field, with some degree of entanglement between the mass and the field: $$\frac{1}{\sqrt{2}}(\ket{0}\ket{\alpha}+\ket{1}\ket{\alpha^{'}})$$, where the two orthogonal states $\ket{0}$, $\ket{1}$ represent different locations of a single mass,  and $\langle \alpha|\alpha^{'}\rangle$ is different from zero (see the earlier discussion for the exact degree of entanglement between the field and the masses, section 3). When one traces out the qubit representing the mass, the other qubit, representing the field, is in a state that is an equal mixture of $\ket{\alpha}$ and $\ket{\alpha^{'}}$. This effect is analogous to what happens for the EM field, where the field gets entangled with a charge that is superposed across different locations. Hence, while it may be possible to prepare superpositions of spacetime geometry by means of a different method, their possibility is not a direct implication of the detection of GIE. One should add that preparing such superpositions of gravitational fields entails measurements in the $\ket{0}\pm\ket{1}$ basis without spatially interfering the two branches in the same location (because this would lead to the same field at the end). As such, this could be exceptionally challenging in practice.} 

{Interestingly, Christodoulou and Rovelli also argued that the discretisation of time could be proven by observing GIE, \cite{CHRO20}. However, one should be careful to note that in fact all observables (classical or quantum) are discrete (to allow for error-correction, see e.g. {\cite{vonNeumann}}), so that discreteness of a variable is not necessarily a proof of its non-classicality. {For instance, a classical bit is discrete, but does not have any non-classical properties. The strength of the GWT and of the GIE experiment is that upon observing GIE one can conclude that gravity is non-classical in a stronger, operational sense, rather than going via weaker properties such as discretisation, which are also displayed by classical systems}.  Finally, Bose and collaborators have appealed to theorems from quantum information theory, stating that Local Operations and Classical Communication (LOCC) cannot create entanglement between two quantum systems, in order to support the claim that GIE rules out the possibility that gravity is a classical decohered channel, \cite{BOMAMO}. {That theorem holds only once one has already assumed that gravity obeys quantum theory, and it allows one to rule out classicalised, decohered versions of it, such as \cite{OPP}. Hence this way of interpreting the detection of GIE is still quite restrictive, albeit more general than ruling out one particular theory of classical gravity versus linear quantum gravity. Moreover, it could be argued that without assuming that the Newtonian gravity is an appropriate limit of some quantum theory, the formula  represents a non-local action-at-distance interaction, thus violating  the basic premise of the LOCC paradigm.}

\subsection{The General Witness Theorem} 

Interestingly, as we mentioned in the introduction, it possible to prove that GIE is a witness of non-classicality with a general witness theorem (GWT), based on more general assumptions than LOCCs theorems, as illustrated by Marletto and Vedral \cite{MAVE17b, MAVE20}. As we said, in order to apply LOCC results one needs to assume that gravity already obeys quantum theory, which is a restrictive assumption given that many proposals for classical gravity question precisely that assumption. Using the GWT, one can use GIE to rule out a larger class of classical models compared to the more narrow approach discussed in the earlier section, thus making the observation of GIE a more conclusive and robust evidence that gravity is not classical. 
We shall discuss here the theoretical assumptions supporting the GWT, and its implications. 

The first assumption is Einstein's {\sl principle of locality}, or no action at a distance. This principle requires that physical variables of a given subsystem (whether or not directly measurable on that subsystem) cannot be affected by operations acting solely on other subsystems.  Note that by the term 'locality' authors have indicated other concepts in the literature on this topic. For instance, there is the no-signalling property of quantum theory, expressed as stating that the local density operator of a given quantum system is not affected by operations acting on other systems only. {There is also Lorentz-covariance in quantum field theory and relativity, which is a particular form of symmetry of equations of motion that are relativistically compliant. The property of locality as defined above is implied by Lorentz-covariance, as the latter rules out instantaneous action at a distance. Locality also implies no-signalling. However, it is more general than either of those, as it is formulated in a dynamics-independent way.}

The second assumption of the GWT is the principle of {\sl interoperability of information}. It expresses the intuitive property that classical information must be copiable from one system that can contain it to any other such system, irrespective of their physical details. In order to formalise this assumption, one can use the constructor theory of information, {(see the appendix for a summary, and the reference \cite{DEUMA} for the formal details)}. {This is a general, unifying framework where quantum and classical systems can be described without committing to the particular formalism of either classical physics or quantum theory, thus being ideal to handle the case at hand. First, one defines an information medium, as a physical system with a set of disjoint states  that can be copied to perfect accuracy. This `copiable' set, which is called an {\sl information variable}, generalises the idea of a classical observable, i.e., a set of states that are all distinguishable. Then the principle is stated as follows: If ${\bf S_1}$ and ${\bf S_2}$ are information media, respectively with information variables $X_1$ and $X_2$, their composite system ${\bf S_1} \oplus {\bf S_2}$ is an information medium with information variable $X_1\times X_2$, where $\times$ denotes the Cartesian product of sets. This informally requires copy-like operations to be possible from ${\bf S_1}$ to ${\bf S_2}$ and vice-versa -- a property that is usually assumed of all good information-carrying systems, but here it is expressed as an exact requirement on interactions between them.}

These two physical principles are obeyed by the best physical theories proposed so far, both classical and quantum: locality is necessary for the existence of independent subsystems, and both principles are necessary for testability.

The third assumption of the GWT is that the interaction between the quantum probes is {\sl mediated} (by gravity in the specific case). This assumption is independently testable via other experiments involving classical gravity interacting with a quantum probe, (e.g., via the COW experiment). It can also be proven assuming unitarity and quantum field theory, as showed by D. Carney, \cite{CAR}. However these assumptions are very stringent, whereas the GWT appeals to less restrictive ones.

By relying on these three assumptions, one need not assume a specific dynamics for gravity in order to demonstrate the GWT. This is essential in order to interpret the test of non-classicality in gravity as ruling out, in case of observing GIE, a large class of classical theories for gravity, not just specific classical models. 

{We shall outline the proof of the GWT in the appendix C. The logic of the proof can be summarised thus: there are two systems (the quantum probes) that are capable of supporting superpositions and entanglement, and a mediating channel. This channel must have the capacity to transfer the non-classicality of these two systems from one to the other, and back again, otherwise it would not be able to entangle them. To do that, it must be non-classical itself. For an interesting quantitative analysis of the non-classicality of the channel within quantum theory, see \cite{PAT24}.}

{Notably, the proof relies on a notion of non-classicality that is more general than any specific dynamics-based definition. It is defined as a lesser property than obeying fully quantum theory. This notion of non-classicality generalises the property of quantum complementarity (i.e. that the variables in questions do not commute) to the case where the non-classical system may not have the full information-processing power as a quantum system. {In a non-classical system there must be two distinct physical variables with the property that they are not jointly distinguishable by the same measuring device, thus satisfying a generalised Heisenberg uncertainty principle. One variable is the classical observable of the physical system at hand - for instance, in the case of the quantum mass, its position. The additional non-classical variable is necessary to enable the non-classical system to mediate entanglement between two other quantum systems. However, unlike the classical variable, it may not be a fully-fledged observable in the sense that its states may not be physically preparable or perfectly distinguishable.}}

The fact that the GWT is dynamics-independent is appealing as it makes GIE detection a more robust piece of evidence to rule out classical models of gravity. It is also appealing that, aside from the assumption of there being some topology on the set of states of the systems in question, the theorem does not assume the presence of a probabilistic structure on the systems in question. Hence The GWT is proven in a setting that is more general than other frameworks, such as Generalised Probabilistic Theories (GPTs), \cite{PLAV}, where probabilities are assumed. So when observing entanglement between the probes, one can rule out all classical theories of gravity obeying the above-mentioned general principles: classical theories that are known and those that could eventually be formulated. 

Since the original proposal \cite{MAVE17b, MAVE20}, other authors have re-cast this theorem in other frameworks. For example, it is possible to propose a no-go theorem using GPTs. In this framework one assumes different axioms, such as no-signalling, the existence of probabilities, and unitarity. This GPT theorem is more general than quantum theory per se, but it has limitations because it relies on a phenomenological description (in terms of probabilities of measurement outcomes), and hence it cannot rule out frameworks that reproduce the experiment with a non-local theory that still satisfies no-signalling, such as Hall and Reginatto's hybrid quantum-classical model, \cite{HARE08, HARE18}, while the constructor theory witness argument can rule those models out. More generally, the main limitation of measurement-based theories such as GPTs is that they cannot model dynamics and do not have a specific ontology, and hence fail to capture key features of quantum theory's (such as its locality, the existence of the Heisenberg picture, and so on). 

Paterek et al. have also proposed a related argument, based on more restrictive dynamics-dependent assumptions, including the unitary formalism of quantum theory. However it is very general within quantum theory as it does not depend on a specific Hamiltonian. Once more, such arguments are dynamics-dependent and hence similar in spirit to LOCCs theorems, \cite{PAT17}. 

Loopholes can always be found, as pointed out by A. Kent et al, just like for Bell-like experiments, \cite{KENT}. Indeed, it is of the essence to be able to independently argue or test the additional principles on which the argument rests, otherwise observing GIE may not prove that gravity is quantum. Moreover, there are classical models which could describe a GIE detection, by violating the assumptions. For instance, as we shall mention, models where gravity is not mediated (see Wheeler-Feynman absorber theory, \cite{FEWH}); and non-local models, such as Hall and Reginatto's. 

Finally, there are other arguments that are based on single-probe experimental schemes, which we shall discuss in section 5.  These arguments however are less general than the GWT in their current formulations, as they assume unitarity or other specific features of quantum theory. %

\subsubsection{Theories that GIE rules out}

Using the GWT, outlined in the previous section, we can now recall what known classical and semi-classical theories of gravity GIE rules out, in addition to general relativity. 

Observing GIE would refute all those theories which regard a semiclassical description of gravity as fundamental, such as quantum field theory in curved spacetime \cite{BIRDAV}. In these theories, the background spacetime is classical, but the back-action of the masses prepared in some quantum state on the field can be taken into account as an average of the energy-momentum tensor in the quantum state of the masses. {In particular, a semiclassical theory of gravity is usually one of these two possibilities: a) it is a limit (taken under some consistency assumptions) of a linear  quantum gravity; b) it is a mean field theory, and as such it preserves a purity of a single quantum system that interacts with the classical one, and thus cannot model the GIE}.  In particular, in quantum field theory in curved spacetime, the Einstein's equation reads: $R_{\mu\nu}-\frac{1}{2}Rg_{\mu \nu}=8\pi G\langle T_{\mu\nu}\rangle$, 
where $R_{\mu\nu}$ is the Ricci tensor; $R$ is its trace; and $g_{\mu \nu}$ is the metric tensor.
These theories provide powerful predictions such as the Unruh effect and the Hawking radiation, \cite{BIRDAV}. Yet, they cannot adequately describe quantum effects in gravity at the fundamental level, as already demonstrated by Page and Geilker's arguments, \cite{PAGGEI}. This is because they resort to a field which is classical - in the sense that it has no pair of non-commuting observables. The field strength at each point is determined by an average of the stress-energy tensor in the quantum state of the masses. In the situation of the GIE experiment, each mass would therefore be affected by the average of the gravitational field generated by the other superposed mass. Supposing each mass is initially in an equally-weighted superposition of the two possible locations, each mass would experience the potential generated by the other mass $m$ positioned at a distance which is the average of the position of the other mass in its quantum state. Assuming once more that only the gravitational interaction affects the branch corresponding to the arm of the interferometer closer to the other (labelled by $1$), the state of the mass would become $\frac{1}{\sqrt{2}}(|0\rangle+ e^{i\phi_m}|1\rangle )$, where $\phi_m=G\frac{m^2t}{\hbar d_m}$ where $d_m=\frac{d_1+d_2}{2}$, and $d_1$ and $d_2$ are, respectively, the distances of path 1 of one interferometer from paths 0 and 1 of the other. Likewise, by symmetry, for the other mass. Thus the phase acquired would be a local phase, which cannot generate entanglement between the masses. Each mass would be undergoing a separate, COW-type experiment: the state of the two masses would be at any time a product state. Semiclassical theories would therefore be refuted as fundamental descriptions of gravity by observing entanglement in the proposed experiment. The same prediction of no entanglement would be reached by models that resort the (non-equivalent) procedure of averaging the linearised quantum gravity Hamiltonian in the quantum state of the two masses. This would also provide only local phases (albeit different from the former case).

Another class of theories that would be ruled out by witnessing GIE are collapse-type models predicting a collapse of the mass wave-function at the experiment's scales, i.e. an irreversible transition to a state where the position is sharp; likewise, similar collapse-inducing hybrid stochastic quantum-classical models, such as \cite{OPP, KATAMI}, would also be refuted. Consider for example the decoherence time a mass of $10^{-12}$ kg superposed across two different locations, approximately $10^{-4}$m apart, (the spatial extent of each interferometer). According to Penrose's collapse models, \cite{PEN, Penrose}, this time is of the order $t=\frac{\hbar}{G\frac{m^2}{d}}\approx 10^{-13}s$, well below $10^{-6}$s, required for our experiment.  {There is a subtle difference between collapse occurring, and decoherence happening while an interference experiment is taking place (the so-called ``fake decoherence" \cite{LEG} or ``false loss of coherence", \cite{UNR}).} For example, as analysed in \cite{LEG}, in neutron interferometry, a neutron spin couples to neighbouring spins and affects their state. Since the neutron is in a superposition of two spatial locations, the neighbouring spins are entangled with the spatial states of the neutron. However, when the two arms of the interferometer are recombined to measure the interference the two environmental states effectively become the same. This is why, despite the neutron having been entangled with environmental spins inside the interferometer, interference is still observable. A massive superposition can also become entangled with the gravitational field (as we explained when we discussed the re-interpretation of Planck's mass) and still evolve coherently. This decoherence could be discriminated from the genuine collapse. {We can also discriminate de-phasing due to gravity from the gravitationally induced collapse, because the former would permit generating entanglement, but the latter would not. In other words, in the case of genuine collapse, given that the dynamics is irreversible, there would not be entanglement generation. While if the dynamics is reversible, entanglement is in principle allowed.} However, we cannot discriminate gravitationally induced spontaneous emission (still a fully quantum effect due to the vacuum state of the gravitational field, but effectively irreversible in practice) from a collapse, as in both cases interference would not occur.

{Thanks to the GWT, we can use the detection of GIE to rule out a number of other quantum-classical models too. There is a rich literature on hybrid quantum-classical systems, see e.g. \cite{BAR} and references therein.  Broadly speaking, they consist of inequivalent dynamical models where a classical and quantum sector coexist and can interact with each other, under given constraints that are required for consistency, which take the form of superselection rules. 
Some of these models are not expressible as the fully decohered version of a quantum dynamical model; hence, it is of the essence to rely on the general theorem in \cite{MAVE20} in order to rule them out as viable models for the generation of entanglement by a classical theory of gravity. 
One of them is the classical-quantum ensemble hybrid model. In particular, Hall and Reginatto \cite{HARE18} asserted that there can be hybrid statistical quantum-classical models where the mediator {\bf M}, despite being classical, can still create entanglement between two quantum probes. 
They analysed a hybrid system made of two quantum probes $Q$, $Q'$ and a classical mediator $C$, using a hybrid model defined as follows. 
{Any point in configuration space is described by the triplet $z=(q,q',x)$, where each real number represents respectively the position of  $Q$,  $Q'$ and $C$.  If $\psi$ is the quantum state of the composite system of $Q$,  $Q'$ and $C$, one can define the dynamical variables $P(z)$ and its conjugate density $S(z)$, with the property that $\psi = P^{\frac{1}{2}}\exp{\left(-i\frac{S}{\hbar}\right)}$.
Then given any quantum observable $\hat M$ and any classical observable $f$, one can define the hybrid ensemble observables as:
\begin{equation}\begin{aligned}
f(x,\nabla_x S) &\to C_f[P,S] := \int dz P f(x,\nabla_x S) \\
\hat M &\to Q_{\hat M}[P,S] := \langle \psi |\hat M |\psi \rangle\;. \label{ENSOB}
\end{aligned}\end{equation}
The dynamics is ruled by the hybrid Poisson brackets
$$ \{ A, B \}_H := \int dz  \Big [  \frac{\delta A}{\delta P} \frac{\delta B}{\delta S} - \frac{\delta A}{\delta S}\frac{\delta B}{\delta P} \Big ] $$
where $\frac{\delta A}{\delta f}$ denotes the variational derivative of the functional $A[f]$ with respect to $f$. This sets a correspondence between the algebra of the quantum and classical sectors to the Poisson algebra of variables $C_f$ and $Q_{\hat M}$ in the configuration space, defined through these relations:
$$
\begin{aligned}
\{ C_f, C_g\}_H = C_{\{ f,g\}_P}\; , \; \{ Q_{\hat M}, Q_{\hat N}\}_H = Q_{[\hat M, \hat N]/i\hbar}
\end{aligned}
$$
with $f,g$ and $\hat M, \hat N$ being respectively any two classical and quantum observables. A classical observable in this context is defined as a functional depending only on $C$'s position $x$ and its classical momentum $\nabla_x S$. Consider the Hamiltonian}
\begin{equation}
H[P,S]= g_1 \int dq dq' dx P(\partial_q S) x +g_2 \int dq dq' dx P(\partial_x S) q'\; \label{EqHam}
\end{equation}
which represents a {\sl pairwise} coupling between $Q$ and $C$ and then between $C$ and $Q'$, {with $g_1$ and $g_2$ being the respective coupling constants.}
By applying the statistical hybrid model dynamical laws one obtains that, for this particular Hamiltonian, the global state of the hybrid ensemble evolves as:
\begin{equation} 
\psi_t (q,q',x) = \sqrt{P(q,q',x)}e^{iS(q,q',x)/\hbar}= e^{ -it H_{eff} /\hbar}\psi_0(q,q',x), \label{EQ1}
\end{equation}
where $H_{eff}=(g_1\hat p\hat x+g_2 \hat q \hat k)$,  $\hat p=-i\hbar \partial_q$ and $\hat k={-i\hbar}\partial_x$ are the momenta operators of $Q$ and $C$ respectively. 
It is easy to check that by starting the three system in say a product state $\psi_0(q,q',x)= \psi_Q(q)\psi_{Q'}(q')\psi_C(x)$, it is possible to obtain a state $\psi_t (q,q',x) $ where $Q$ and $Q'$ are entangled, by evolving it as specified by $H$. 
As explained by Marletto and E. Marconato, \cite{MAMA}, the proposed model in fact validates the witness. In short, this is because the model conceals a hidden non-locality in the configuration space dynamics, in contradiction with the locality assumption of the GWT. {Indeed, such statistical hybrid is non-linear and moreover the classical and quantum sectors are not fully independent, due to the fact that the Poisson brackets of a variable of the classical sector with a variable from the quantum sector is non-zero.} Moreover, even when considering a local account, the model involves a non-classical mediator, with non-commuting variables $x$ and $\hat k$ that can be empirically accessed, as required by the witness. Likewise, recently proposed models such as \cite{PLAV, KHONIM} all resort to the classical Newtonian potential and can generate entanglement, but violate the assumption of locality, thus corroborating the witness of non-classicality. 

Other models that can be ruled out on the account of the witness are the hybrid quantum-classical systems proposed by T. Sherry and E. Sudarshan, \cite{SUD}. These models are based on a superselection rule, whereby a quantum system is coupled via a standard Hamiltonian with another system that has a non-classical variable which by the superselection rule is required to be unobservable. It is very interesting that when using these hybrids to model GIE, one can generate entanglement between the two masses, but this violates the superselection rule. This is because the generation of entanglement is equivalent to indirectly measuring the superselected variable. Hence these kinds of hybrids, if they are able to generate entanglement, are in fact fully non-classical as per criterion of section.}

{It was suggested that the GIE is not sufficient to test the quantum nature of gravity because it probes only gauge degrees of freedom, \cite{ANHU}. Rovelli and collaborators argued against this point of view, by stating that the non-classicality uncovered in one gauge is gauge-independent, and must exist in other gauges too. \cite{CHRO19}. Bose at al. and Marletto and Vedral pointed out how the dynamical perturbation induced by the quantum probe cannot be gauged away, and that the GIE effect is not a completely static one, but it is in fact dynamical, hence it is not pure gauge as the static Newtonian/Coulomb field regime would suggest, \cite{MAVEQUE}. In \cite{MAVE23}, Marletto and Vedral claimed that detecting entanglement is equivalent to measuring scalar modes of the EM field, and by analogy of the gravitational field; this effect shows that these modes are physical. This study suggests that also in other gauges the same should be true, even if these modes are not directly detectable.}

{As we mentioned, there exist theories that do not assume a mediator for the interaction between charges or masses - hence if there is no mediator, there is nothing to quantise. In that sense, the non-classicality witness we discussed does not apply to them. Carney \cite{CAR} has argued that assuming unitarity and conservation of energy-momentum it is possible to show that a mediator for gravity is necessary -- these assumptions are quite restrictive, and it would be desirable to have other reasons to reject non-mediated theories of gravity. {One of the reasons is that classical gravity (as described by general relativity) has been confirmed experimentally to be mediated, for example via the detection of the finite speed of propagation in gravitational waves, \cite{gwaves}}. However there are some theories, such as those by Hoyle and Narlikar \cite{HONA}, that do away with the mediator by imposing certain special boundary conditions and accepting action at a distance of dynamical variables that are not directly observables. Like other proposals, this theory is ruled out a-priori by the principle of locality. Such theories also have to modify the principles of energy and momentum conservation to make up for the absence of the field backaction (see arguments by van Dam and Wigner, \cite{WIG1}). It can also be speculated that if one applied the GWT to a local version of such a theory, it would imply that the boundary conditions must be quantised to describe the generation of entanglement.}

\subsection{What would it mean not to observe GIE?}

Failing to observe GIE could mean a number of interesting possibilities, which makes the GIE experiment even more promising. Given the natural asymmetry of the witness (GIE is only a sufficient, but not necessary condition for non-classicality), failing to observe GIE does not demonstrate that gravity is classical. One of the conclusions consistent with no GIE is certainly that gravity is classical after all, or that a gravitational collapse occurs at the relevant scale. However, it could be that some of the assumptions of the witness are false (particularly, locality or interoperability of information). It could also be that gravity is quantum, but obeys a different model of quantum gravity, and hence the magnitude of GIE calculated with LQG is incorrect; for instance, one could have a model that is non-linear (see for instance \cite{STA1, STA2, CARREV2}). Then there are even more remote possibilities: such as other fields being at play (the fifth force?), which are relevant at the scale of the experiment. All these would be very exciting avenues and one would need to look at how to develop further experiments to be able to discriminate between them. 

More generally, {the fact that the phase between different branches of a spatially superposed mass is decohered via gravity} does not necessarily indicate that gravity is classical. In the next subsection, we shall discuss in more detail why gravitational decoherence is not evidence of classicality of gravity.

\subsubsection{Gravitational decoherence is not sufficient evidence for the classicality of gravity} 

Gravitational decoherence and collapse, which may cause null entanglement in a GIE experiment, could simply be just like any other form of decoherence that occurs under unitary quantum theory, \cite{SCH, ZUR}. Namely, their origin could be due to the entanglement created between the system and the quantised states of the gravitational field (assuming the linear regime) - hence, gravitational decoherence per se is not a proof that gravity is classical.  The mechanism for unitary decoherence is simple. We assume that a mass $m$ is superposed across a distance $\Delta x$ in some background gravitational field ({Note that one could have a superposition of different values of energy, see the discussion in section 5.A}. The two different locations will lead to different interactions with the underlying gravitational field. The field will initially be assumed to be in a coherent state as is appropriate for the classical regime. ({One could also use a mixture of coherent states, such as a thermal state, in which case the same analysis would work, using linearity}). If the mass is in one location, it displaces the gravitational coherent state one way, while if it is in the other location, it displace the field differently. The amount of displacement will be proportional to the ratio of the energy difference between two locations of the particle to the Planck energy, $\Delta E/E_P$, where $E_P=m_Pc^2$ and $m_P$ is the Planck mass. This way, the two massive spatial states become entangled to the gravitational field, which, when the field is traced out, leads to decoherence (just like any other type of decoherence in quantum physics). 

It is, however, also conceivable, that the loss of coherence is due to the interaction with the classical gravitational field (though semi-classical models suffer from many inconsistencies which we will ignore in the present discussion). In this case, the field affects the phase between the two massive states, but does not become entangled to them (since it is now assumed to be classical and therefore incapable of exhibiting entanglement). This kind of noise is sometimes referred to as dephasing. In the appendix A, we show how to discriminate this form of ``classical" dephasing from the genuine quantum decoherence. In summary, the answer will be via the effect called ``spin echo": classical dephasing can always, at least in principle, be reversed by acting only on the system; the decoherence due to the entanglement with the environment needs us to be able to act on the environment too, or to actively error-correct to keep the system disentangled from the environment. 

Neither entanglement-induced decoherence, nor classical dephasing are what is traditionally thought of as the objective collapse, \cite{PEN} - in which gravity somehow forces a modification of quantum physics by leading to collapse, but not by either of the two (unitarity-compatible) methods described 
above. This objective collapse too can in principle be discriminated from other forms of decoherence. Interestingly, however, and as far as the dynamics of the system is concerned, the gravitational collapse can always be thought of as a form of decoherence. This is perhaps not surprising given that any non-unitary, but completely positive, quantum evolution can be viewed as a unitary one from the perspective of the higher Hilbert space (assuming that collapse is a completely positive transformation). To discriminate a genuine collapse from this higher level decoherence processes would then require us to have the full knowledge and control of the environment in which case the relevant experiment is just an interference experiment that also involves the environment. While this may be hard in practice, it is certainly at least always possible in principle. 

{The literature on this topic is vast, but frequently confusing and contradictory. In our view this mainly reflects the nature of the topic and the present lack of experimental evidence regarding quantum effects in the gravitational field. {Here we shall discuss a model inspired by the following references: \cite{Boughn,Weinberg,Dyson,Kay,Blencowe,CJR,MAVE18,BOMAMO,Skagerstam, LIN1, LIN2}}. It is worth bearing in mind that, albeit differently motivated, they all have in common the linear model of gravity. The most relevant paper that addresses the classical gravitational dephasing is \cite{Brukner} and the literature on genuine collapse can be found here \cite{Karolyhazy,Diosi,Penrose}. The fact that semiclassical treatment of gravity is physically inconsistent was first clearly spelt out in \cite{PAGGEI} (see also \cite{Ford}), but we will ignore this fact and assume that gravity could in principle still be treated classically when coupling to quantum states of matter.} 

Finally, {we recall the phenomenon of ``fake decoherence", \cite{LEG} (also called ``false loss of coherence", \cite{UNR})}, where the gravitational entanglement to the massive superposition does not actually prevent us from observing interference. This is perhaps the most subtle message here, namely that not everything that looks like decoherence, actually leads to ``irreversible", or ``true", decoherence.{When a neutron undergoes interference as in \cite{COW}, its spin couples to the neighbouring spins. Since in the interferometer the neutron exists in a superposition of 
two different spatial locations, the neighbouring spins are in two different states which are entangled with the spatial states of the neutron. If we were to trace out the environmental spins, the neutron spatial superposition would seemingly decohere. This however does not prevent us from observing neutron interference! It is so because when the two arms of the interferometer are recombined to measure the interference, the two environmental states also merge into effectively the same state. This is why despite the fact that the neutron was fully entangled with environmental spins inside the interferometer we are still able to observe interference at the end. The same applies to the GIE experiment, where the reversibly-created entanglement with the field is necessary to create the entanglement with the masses, but does not lead to decoherence of the joint state of the two quantum masses'. }

{The same would occur with a massive superposition and the gravitational field: under the assumption of unitary quantum dynamics, gravity is responsible for true decoherence (meaning an irreversible effect) only if the two states of the gravitational field coupling to the mass do not return back to the same state when the two massive states are recombined to interfere. Therefore, a massive superposition does necessarily not collapse due to interaction with gravity; what would lead to true decoherence, for instance, is an emission of a graviton which would for all practical purposes be irreversible (just like a spontaneous emission of a photon); or an interaction with no graviton emission, but capable of propagating the phase information to an arbitrarily large distance, see e.g.\cite{LIN1, Blencowe}.}  

This naturally leads us to revisit the scenario in \cite{BAOZ} where a massive particle undergoing a double slit interference interacts with another massive particle such that the resulting configuration reveals which slit the particle has gone through (thereby destroying interference). The conclusion of that paper is that in order for classical gravity not to spoil interference, the interference fringes would have to be smaller than the Planck length. This result too, can be modelled with the fully quantised field. Briefly, the logic is as follows. The energy difference between the two configurations where the mass is at the two slits should be at least as big as a single graviton energy (whose wavelength should not be smaller than the distance between the slits in order not to destroy the interference). This graviton would be emitted by the process of gravitational bremsstrahlung \cite{CJR}. Therefore, $Gm^2 d/r^2 \geq hc/d$, where $r$ is the distance between the two masses and $d (<<r)$ is the slit separation. This is exactly the condition in the paper \cite{BAOZ} that leads the authors to conclude that the size of the interference fringes would be below the Planck length (and therefore presumably unobservable). 

The fact that a quantum gravitational argument could be used to reach the same conclusion reinforces the idea that any classical collapse (as in \cite{BAOZ} or \cite{Brukner}) can be reproduced quantum mechanically. We therefore have to be careful regarding our conclusions about the quantum nature of gravity just based on the bare fact that a certain experiment is (or isn't) capable of demonstrating any quantum interference effects. After all, in a fully quantum universe, the classical world exist only because of the entanglements between various subsystems whose quantum interference effects are therefore impeded due to the overall higher level of quantumness  \cite{Joos}. Even the classical world is classical only because the universe is (at least to a high enough degree) quantum.

\section{Variants of the GIE experiment} 

The GIE experiment is based on creating gravitational entanglement between two different probes, in a setup that involves two spatial superpositions. However, many variants are possible because one can: 1) Vary the degrees of freedom that get entangled: not just position degrees of freedom, but energy, or frequency, or mass; 2) Change the configuration of the quantum probes, to to amplify the GIE effect; 3) Vary the assumptions the experiment is based on, to reduce its complexity (prominently single mass variants). We shall briefly discuss these possibilities.

\subsection{GIE with alternative degrees of freedom}

The strategy of looking for GIE between different degrees of freedom (different from position) is followed by a few proposals that have been put forward soon after the seminal papers. 
In Deutsch, Vedral and Marletto, \cite{MAVEDE}, an interesting observation is made, that it is possible to entangle gravitationally two particles that are superposed across different values of their masses, rather than of their positions. A numerical analysis shows that it would be possible to detect GIE in an experiment using two hypothetical neutrino-like particles with a mass much larger than neutrinos. 
{The relevant observables here are the energy of the neutrino-like particle, whose eigenstates $\ket{m_i}$ are each labelled by a particular value of the mass, $m_i$. The other (complementary) observable is the equivalent of the neutrino's flavour, with three eigenstates $\ket{\nu_i}$.  
When a neutron decays into a proton and an electron, a neutrino is produced, in an eigenstate of its flavour. Here we can suppose a neutrino-like particle is similarly spontaneously emitted in a flavour eigenstate $\nu_1$.}
{The flavour eigenstate can be written as a superposition of the energy eigenstates:
\begin{equation}
\ket{\nu_1}=\cos (\theta)\ket{m_1}+\sin(\theta)\ket{m_2}
\end{equation}
where $\theta$ is the mixing angle -- a fixed parameter that depends on the physics of weak interactions.}
{The states of mass superposition would have to be achieved by a natural decaying process, whereby the mass superselection rule is satisfied overall.}
{The emitted particle has a Hamiltonian $H$ that leads to its free evolution, so that the state at time $t$ reads:
\begin{equation}
\ket{\psi(t)}=\cos(\theta)\exp\left(\frac{E_1}{\hbar}t\right)\ket{m_1}+\sin(\theta)\exp\left(\frac{E_2}{\hbar}t\right)\ket{m_2}\;,
\end{equation}
where $H$'s eigenvalue in the state $m_i$ is $E_i^2=m_i^2c^4+p_i^2c^2$.}

{Imagine two neutrino-like particles $N_1$ and $N_2$ travelling along parallel paths, whose distance is $d$. Supposing that initially they are both in the same quantum superposition of two masses $m_1$ and $m_2$, specifically the flavour eigenstate $\ket{\nu_1}$, the initial state of the two neutrino-like particles $\ket{\nu_1}\ket{\nu_1}$ evolves to the state $\ket{\phi(t)}=\sum_{i,j}\alpha_{ij}\ket{m_i m_j}$ where the phases are defined as follows:
$$\alpha_{ij}=\exp{\left(\frac{E_i+E_j}{\hbar}+\frac{Gm_im_j}{\hbar d}\right)\Delta t}\;.$$ This state describes two neutrino-like particles that are entangled to a degree that varies with time, according to the magnitude of the phase. While with the neutrino's masses the GIE would be far too small to be detectable, masses close to Planck's would provide a detectable GIE. Indeed there are speculations that the Planck mass could correspond to the mass of the largest fundamental particle. The hope of this proposal is to use neutrino-like oscillations that occur naturally once the particles are emitted to detect the GIE signature. How to detect such particles is an open question, but the main point of this work is to explain that one can generate GIE between different degrees of freedom from the position of the masses.} 

{In Vedral and Marletto, \cite{MAVESAG}, a matter Sagnac interferometer with a single particle of mass $m$ is used to probe the quantum nature of the gravitational field. The argument has, as additional assumption, the quantum equivalence principle, as defined in \cite{MAVE20a}; remarkably, there is no actual gravity present. Here the GIE is created by a particle in an interferometer, superposed across two different distances from the center of the disk; and by the disk spinning in a superposition of two angular frequencies. The degrees of freedom that give rise to the entanglement are the positions of the particle and the two different angular momenta of the particle.} 
The phase that is responsible for the effect reads: $$\phi_S= \frac{2m}{\hbar}\Omega A$$ where $\Omega$ is the angular frequency of the particle and $A$ is the area {enclosed by the particle spinning on the disc}. Entanglement is created between different degrees of freedom of the same massive particle, the position and the angular momentum values. 
{It is worth mentioning two more implementations (which we shall discuss in section 6): freely falling masses and nanomechanical oscillators. The former proposal simply capitalises on the fact that each mass undergoes the process of quantum wavepacket spreading in position, and gravity acts to entangle them. With nanomechanical oscillators it is their vibrational degrees that provide the spatial extent of each oscillator and again the gravity acts accordingly. For instance, in  \cite{KRI}, two identical massive spheres are trapped in two harmonic potentials, cooled down close to their ground state, and let free to interact, thereby generating GIE by entangling their position degrees of freedom.}  

\subsection{Alternative configurations}

Given that gravity acts on anything with energy and is likewise acted upon by anything having energy, one can imagine many different ways of performing the GIE tests. This is encouraging given that some of these implementations could be easier to realise experimentally. Some recent proposals have used techniques of quantum entanglement detection theory to improve the sensitivity of the GIE original setup. In \cite{FEFE2},  a novel method has been proposed by T. Feng and Vedral to increase the sensitivity of weak entanglement detection by several orders of magnitude.  {This relies on the fact that the so-called weak value can arbitrarily exceed the eigenvalues of the conventional observable, however, to measure such weak value one always relies on post-selection. In that sense, the gravitationally induced entanglement can be amplified, but at the cost of increasing the number of trials in order to implement post-selection (see \cite{RUG})}. Moreover, there have been proposals to generalise this idea using three probes instead of two -- it has been found that three probes can lead to a more stable setup against decoherence, \cite{SCHUT, TILLY}. J. S. Pedernales and collaborators, \cite{Pern} have also suggested a method whereby a massive body can be used to amplify the gravitational interaction between two test probes. 

\subsection{Single-probe variants}

The second strategy inspires single-probe variants of the experiment, which have been proposed in the hope that they could make the experiment easier. Dealing only with one quantum probe, these variants present a lesser challenge than the GIE experiment, but they need additional, stronger theoretical assumptions to work as witnesses of non-classicality, hence losing in generality compared to the GIE experiment. 

For instance, R. Howl and collaborators \cite{HOVENA} have proposed an interesting idea based on the generation of non-Gaussianity in the quantum state of a Bose-Einstein condensate, as a signature of quantum effects in gravity. The self-gravitational interaction of a Bose-Einstein condensate can be treated both as a classical system and as a quantum system, leading to two different Hamiltonians: the quantum Hamiltonian is capable of generating non-Gaussianity in the condensate, while its classical version cannot generate non-Gaussianity.  In particular, when we have a term proportional to the square of the density of particles $n^2$ (as an operator), then in the semiclassical treatment of gravity, this self-energy could be written as $\langle n\rangle n$. In other words, if gravity is classical, the quadratic interaction is identical to what could be called ``a mean-field theory" treatment.  {Note that this argument is not based on a causally propagating mediator with a local interaction as it is simply a single mass acting on itself at one and the same point.}
One can then experimentally test which of the two Hamiltonians is corroborated in a test involving a Bose-Einstein condensate at the laboratory scale: if non-Gaussianity is generated, the classical model for gravity is ruled out since the linear Hamiltonian $\langle n\rangle n$ is not capable of generating non-Gaussian states. However, other classical models for gravity could still be viable, hence this test has a much narrower domain of applicability than the GIE experiment. 

In \cite{MANAYA}, A. Matsumura and collaborators applied Leggett-Garg inequalities to the evolution of a single mass interacting with an harmonic oscillator representing the gravitational field, in order to conclude the non-classicality of gravity if the equalities are violated. 
 In \cite{BEWAGI}, A. Belenchia and collaborators proposed that an experiment utilising a single mass or charge in a quantum superposition, interacting with another test charge/mass, is sufficient to conclude for the quantisation of gravity, assuming unitarity and causality.  Once more, both these proposals require the assumption of overall unitary evolution, hence being quite narrow in scope. {Moreover, The oscillator produces the classical gravitation potential. In this sense the proposal is not different from the EM case without the EM waves.}

More generally, one can treat a single mass in the second-quantised picture, where the creation of a superposition of different spatial locations corresponds to creating an entangled pair of modes, \cite{DUNN}. In this sense, assuming that the generation of this superposition is mediated by gravity, one can apply the non-classicality witness that underlies GIE, and obtain the conclusion that gravity is quantum. 

It is worth repeating that an experiment with only one massive superposition acquiring a gravitational phase shift (as in Feynman's original suggestion) cannot tell us anything about the quantum nature of gravity. Either successfully interfering a massive particle, or experimentally demonstrating that this cannot be done, is compatible with both possibilities, that gravity is classical or indeed quantum. The reason for this is as follows.  A typical interference experiment involves starting with a mass localised at some point $x_1$, then creating a spatial superposition say across two locations $x_1$ and $x_2$ and then finally re-interfering these two possibilities back at say $x_1$. Now suppose that we do not observe any interference (loosely speaking the probability would be there for the particle to be detected at $x_2$ as well). This could be because gravity collapsed the quantum superposition, however, both the classical gravity and quantum gravity could act as such noise. So if we are only able to measure the massive particle (and do not also probe the field in the same experiment) the failure of quantum interference does not tell us whether gravity is quantum or classical. (See the discussion about decoherence in section 4). 
The same goes for the case if we observe interference. In this case, gravity could be classical and simply not able to dephase the superposition within the time frame of the experiment (it is too slow, yet classical). Also, it could entangle to the mass, say in the way that the state becomes $|x_1\rangle |g_1\rangle + |x_2\rangle |g_2\rangle$, where $|g_1\rangle$ and $|g_1\rangle$ are two, possibly even orthogonal, states of the gravitational field. One might then be tempted to say that the mass is not in a spatial superposition because when the gravitational field is traced out, the resulting state of the mass is a mixture which therefore ought not to be able to interfere. However, this kind of decoherence is only superficial (also known as the false/fake decoherence,\cite{LEG, UNR}). The reason for this is that when the two massive states are brought back to the same location to interfere, the two states of the field also merge back into one and the same state. The gravitational field in this case, even though it is quantum and therefore capable of entangling itself to the mass, still enables it to ultimately be able to interfere. So both classical and quantum gravity are in principle compatible with massive superpositions or their collapse.

\subsection{Towards a general argument in support of the single-probe variants}

It is possible however to use extra assumptions in order to conclude that even a single-mass experiment could reveal quantum effects in gravity. This could result in a `temporal' version of the GWT, where it is the dynamical (temporal) evolution of a single probe, induced by gravity, that reveals gravity's quantum features.

G. Di Pietra and Marletto, \cite{DIMA}, have proposed to use the conservation of an additive quantity (e.g. energy) together with the formalism of quantum theory in order to provide a special case of the temporal version of the entanglement-based GWT. In this `temporal' version of the GWT, the mediator causes the coherent evolution of a quantum system, under a Completely Positive Trace-Preserving (CPTP) map that also conserves a global additive observable. If the mediator can cause the coherent evolution of the quantum probe, then the argument proposed by the authors allows one to conclude that the mediator is non-classical. (See Appendix C for a summary of the argument)

In \cite{FEFE1}, T. Feng and collaborators have proposed an argument to claim that conservation of momentum implies the quantization of gravity -- not just the exact conservation as in \cite{DIMA}, but also the average conservation. This argument is then applied to analysing the free fall of a particle, which, with stronger, extra assumptions, could be seen as implying the quantisation of gravity.

The arguments just outlined all have the drawback of relying on quantum theory's formalism, so they do not have the full generality of the GWT based on entanglement between two probes and a mediator. 
However they may inform easier single-probe experiments, which may also be important stepping stones towards realising the full gravity experiment. The generality of the GWT suggests that there could be a similar argument for non-classicality based on a single mass, with extra axioms. In \cite{DIMA} the axiom that plays the role of locality is a conservation for an additive quantity. This is suggestive of a possible temporal argument that could serve as a complement to the spatial version of the witness, based on spatial entanglement. Considering the symmetry between time- and space- correlations in quantum theory, one can expect that this generalisation is possible.

\section{Experimental efforts}

GIE can be implemented in multiple platforms and technologies. As discussed in \cite{MAVE17b}, one can think of the general experimental scheme to witness GIE, with two quantum probes becoming entangled via gravity, as a general quantum computing algorithm, that can then be realised in different kinds of qubits (the quantum probes) capable of interacting gravitationally.  {The experimental efforts are still in their infancy, as we shall explain in this section}. However, there are already promising feasibility studies in the literature and a few actual experiments that provide elementary building blocks for realising the actual GIE detection platform. The actual GIE experiment realisation may still be far in the future, but its consequences would be epoch-making. Furthermore, while working towards the goal of constructing the full GIE detecting platform, there are a number of exciting milestones that can be realised, \cite{CARREV, CARREV2}, including a single interferometer with a mass of the order of a nanogram, which itself would pave the way to a novel class of quantum sensors for the gravitational field, and would allow many foundational questions relating to the nature of gravity to be addressed. Here we review the main possible experimental platforms that are currently being investigated in laboratories around the world. The race is on to realise this experiment and the technologies listed here are not exhaustive of the possible platforms that could be used, but appear to be the most promising given our current knowledge. 

{\bf Stern-Gerlach interferometry.} The first promising possibility, proposed by S. Bose and collaborators, is to use two masses with an internal degree of freedom (spin) that is acted upon to create the massive superpositions, \cite{MAMABO}. {In this scheme each quantum probe is acted upon by a Stern-Gerlach interferometer that first creates the spatial superposition by means of a magnetic field gradient that acts on the spin degrees of freedom of the probes; then lets the probes interact gravitationally to create GIE, and then undoes the spatial superposition by means of another magnetic field gradient.} All this can happen only by manipulating the spin degree of freedom of each probe, not their center of mass -- thereby achieving good control of the probes with existing technologies. In particular, one can consider two neutral microcrystals of $m = 10^{-14}$kg, each with an internal spin degree of freedom, that is used to set each crystal into a superposition of the 0 and 1 path, with an extent of $\sim100\mu m$. The probes, once in a superposition, can either be dropped to undergo free fall, \cite{MAMABO}, or still be guided by the spin degree of freedom, while interacting gravitationally for a total time of $1-2$s.   In the seminal proposal discussed in \cite{MAMABO}, the microcrystals (of about 1 $\mu m$ in diameter) are separated by $\sim 450\mu m$ at the point of closest approach. This distance has been calculated to be sufficient to reduce the disturbance caused by the Casimir-Polder effect to a minimum. An alternative way to reduce the Casimir-Polder effect is to encase each interferometer in a Faraday cage and separate them by a conducting plate shield, as carefully explained in \cite{CAS1, CAS2}. This allows one to reduce the distance and the overall superposition extent by about one order of magnitude. 
Remarkably, the Stern-Gerlach device operating in a closed interferometry loop has been recently realised with Bose-Einstein condensate of $10^4$ Rb atoms, \cite{MAMAFO, MARGIT}. These masses that are far lower than those needed for the GIE experiment; nonetheless this has been a significant and encouraging step forward towards realising this interferometer with probes having an appropriate mass to detect GIE.

One possible way to realise the GIE quantum probe in this Stern-Gerlach device is a levitated nanodiamond with a negatively charged nitrogen vacancy (NV-) centre, which carries an electronic spin 1. A magnetic field acts on an NV- spin to prepare the nanodiamond in a spatial superposition. The spatial superposition is then recombined and the spin is measured at the end with an optical redout unit. This setup can be realised first for a single nanodiamond, to detect a gravitational phase difference due to another nanodiamond or to the earth's gravitational field. In \cite{WOGA} the main features and challenges of the NV- centre technology are analysed in detail. The main strength of the approach is the possibility of optical preparation and readout that manipulate directly only the spin, not the center of mass of the nanodiamond. One of the main difficulties with this technique is that the spin coherence time is severely limited in nanodiamonds compared to bulk diamond, with the longest time recorded at room temperature being of the order of 700 $\mu$s. 
The trapped configuration has advantages compared to the free-falling scheme, and is currently the most viable path (see Genovese et al., \cite{GEN}). As explained in \cite{GEN}, one needs to use a magnetic structure with a size of the order of 10 m, to achieve free fall of about 2s in time, and also to achieve control of the particles over the nm scale. Implementing vacuum and cryogenic systems over a structure of this size is very challenging. In the trapped configuration instead there is a more compact setup, and also the ND can be recycled at the end for another experiment, while in the free-falling setup they end up lost at every round.  
 
 {\bf Matter-wave interferometry.} A second interesting possibility to realise the GIE experiment is the use of matter-wave interferometry, \cite{MAVE17a, MAVE17b}. One would use, like in the previously discussed Stern-Gerlach setup, two adjacent interferometers. While this possibility has been briefly discussed by Marletto and Vedral in \cite{MAVE17b}, there is no systematic feasibility study. In \cite{MAVE17b}, superposition extents of $1\mu m$, a time of flight of $1\mu s$ and masses of $10^{-12}$kg were considered. The same considerations regarding the Casimir-Polder potential apply to this method too, so the interferometers would have to be either distanced appropriately, or have some screening in place to shield the effect, \cite{CAS1, CAS2}.  
 
Typically, among the largest quantum probes that are currently usable in matter-wave interferometry, one has atom condensates or complex molecules. M. Arndt and collaborators have described the recent challenges and advances in interferometry of complex molecules (metals and dielectrics), using the so-called Talbot-Lau interferometry, \cite{ARN}. This is a type of near-field interferometry that uses the Talbot-Lau effect, whereby the coherent wave impinging on a grating undergoes diffraction. The diffraction pattern repeats itself periodically at specific distances from the grating. As explained in \cite{ARN}, this method offers a number of important advantages over other types of matter-wave interferometers (such as the Mach-Zehnder interferometry), among which there is the fact that the requirements for collimation are much less stringent. The molecules currently accessible with this kind of interferometry are at present of the order of $10^-21$kg, or $10^6$ amu, which is far below the necessary order of magnitude for GIE detection. 
 
Matter-wave interferometers have been also realised with Bose-Einstein condensates and atom clouds. Bose-Einstein condensates consist usually of $10^4$ atoms, thus being much smaller than the masses required for the GIE experiment. However, they can be useful in realising single mass variants of the experiment, as for instance described by Howl and collaborators \cite{HOVENA} (see section 4). In \cite{KAS}, a team led by M. Kasevich used an atomic Rb cloud to witness the gravitational AB effect. One can measure the gravitational phase shift induced by a kilogram-scale source mass close to one of the wave packets, in analogy with the Aharonov-Bohm phase shift induced by a solenoid on a charged particle. This experiment deals with much smaller masses than those needed for GIE, but establishes a benchmark for the control matter wave-packets.

{\bf Trapped oscillators.} A third, alternative, possibility is to use quantum probes that are initially trapped with some potential and then released, to interact gravitationally, as proposed by T. Krisnanda and collaborators in \cite{PAT18}, see figure 5. 
\begin{figure}[h]
	\centering
	\includegraphics[scale=0.4]{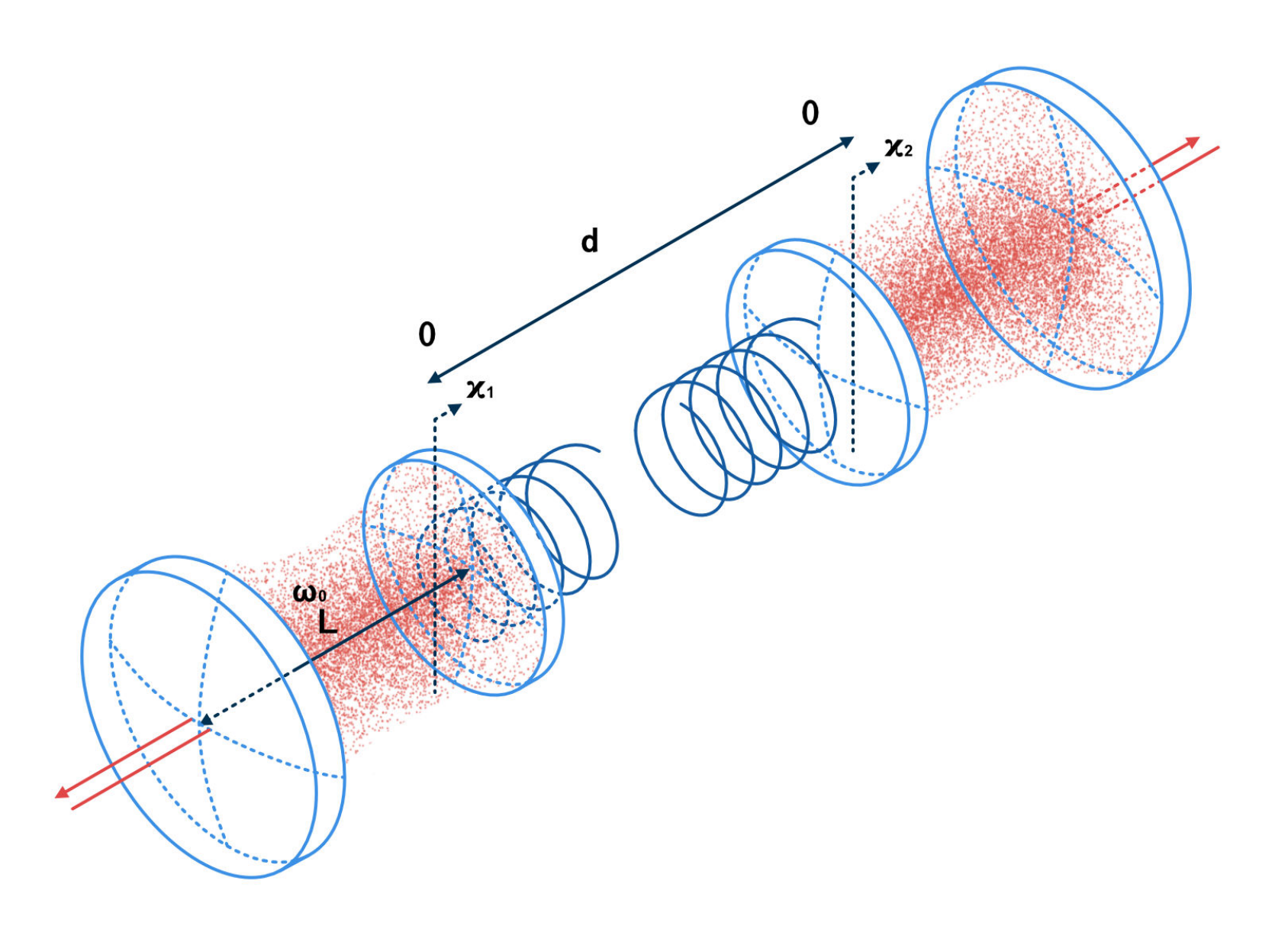} 
	\caption{A schematic representation of nanomechanical oscillators that can interact via gravity to produce GIE. Each oscillator (represented here schematically by a spring) is mounted on a mirror of an optical cavity (of lenght $L$ and characteristic frequency $\omega_0$). The optical cavity is used for the readout. The relevant degree of freedom for the superposition that leads to GIE is the displacement operator $xi$ of each mirror, $i=1,2$. }
\end{figure}

The readout in this case is performed with weakly coupled light fields. Theoretical modelling including decoherence due to air scattering, thermal photons and Casimir forces, as discussed in \cite{PAT18}, shows that this setup requires masses and frequencies in the range of current technological capabilities: for instance a possible configuration uses two Osmium spheres of $100\mu$g, prepared in two Gaussian states, at a distance of $0.3$ mm and then released, oscillating with frequency $0.1Hz$ \cite{PAT18}. This configurations leads to detectable entanglement within seconds, well within the coherence times of the system. In this regime, gravity dominates the Casimir-Polder potential, so the latter leads to negligible contributions. The advantage of this scheme is that it does not require the preparation of a macroscopic superposition; however one needs to use much larger masses (100 micrograms) than with the other two proposal that deploy standard interferometers (where the masses are of the order of a nanogram). 
A realistic possibility to realise this scheme are nanomechanical oscillators, {whose physics is reviewed in e.g. \cite{REVNANO}.} These can be cooled to their ground state with current technologies, \cite{OPTO1, OPTO2}. In \cite{ASPEL}, a team led by M. Aspelmeyer realised a coupled, nanoscale optical and mechanical resonator consisting of a nanobeam formed in a silicon microchip. The resonator is then cooled to its ground state using the radiation pressure from a laser. The cooling is realised at an environmental temperature of about 20 K. Two such resonators, set up in an adjacent configuration, could realise the detection of GIE according to this proposed setup.

{\bf Main challenges.} In general, there are two kinds of challenges for the GIE experiment. One kind is the usual challenge for any experiment that requires to create macroscopic quantum superpositions and maintain them for considerable time: there are several possible sources of decoherence, and they need to be managed. This case is particularly challenging because the gravitational interactions are much weaker than electro-magnetic interactions.  The main sources of decoherence for a GIE experiment are: scattering with molecules of air, scattering with thermal photons, electro-magnetic interactions, blackbody emission and absorption, gradient of gravity, for instance due to imperfect alignment of the interferometers with respect to the Earth's surface; and gravitational de-phasing due to massive objects near the experimental setup (see the analysis in \cite{GUN}).

The other kind of challenge is that one needs to couple the two interferometers gravitationally {\sl only}, and other interactions (e.g. Casimir, Van der Waals, etc.) must either be negligible in the specific regime or clearly distinguishable from the gravitational interaction, so that the phases induced by them can be told apart from the GIE phase. If such interactions were comparable to gravity, it might still be possible to isolate the characteristic $\frac{1}{r}$ behaviour of the gravitational potential in the phase  as opposed to other potentials, see e.g. the analysis by Chevalier and collaborators \cite{CHEV}.

\section{Open problems}

The field of GIE as a test of quantum gravity effects is rapidly expanding, both from the experimental and theoretical point of view. Indeed, there are many open questions that lie ahead of us in a multitude of directions. A few of the most interesting and pressing issues are summarised below.

\begin{itemize}
\item{} We know that all non-perturbative approaches agree with linear quantum gravity in the low energy regime of the GIE experiment, however it would be theoretically important to understand how a non-perturbative model would explain the GIE generation and detection. {One could speculate that the metric may be one of the non-commuting degrees of freedom, but then it would be necessary to identify another degree of freedom that does not commute with it. To the best of our knowledge, there is at present no explicit calculation of the GIE effect in non-perturbative models}. Hence, an interesting open question is finding the degrees of freedom of gravity that mediate GIE in non-perturbative models of quantum gravity, such as loop quantum gravity or string theory.
\item{} We have discussed that the GIE test relies on a general witness of non-classicality, and specifically on the the GWT. This theorem has a temporal counterpart, which albeit less general in its assumptions, informs single-mass tests of quantum gravity. Could one have a unified, general argument that includes both a spatial and temporal argument for the witnesses of non-classicality? If not, what is the reason for that?
\item{} The implications of the GIE experiment for the observables of the gravitational field are also an interesting open problem. It has been suggested \cite{MAVE23} that observing quantum phases on charges is an indirect detection of ghost modes of the electromagnetic field, and that a similar conclusion holds for GIE and the ghost modes of the gravitational field. Extending explicitly this analysis to the gravitational field is important, especially because it may exposes inconsistencies of existing quantisation procedures.
\item{} Currently the most investigated set-up to generate GIE uses an levitated diamond with an NV centre spin as the control degree of freedom,  \cite{BOMAMO}.  However, as we discussed, the GIE experimental scheme is general, so it is not tied to a particular technology. It could be realised with many other methodologies. It would be therefore important to investigate other experimental implementations in the near future, to allow for diversity of approaches. 
\item{} Variants of the GIE detection mechanism may be able to detect higher-order corrections to the quasi-Newtonian regime predicted by non-perturbative models of quantum gravity. It would be important to investigate them, or to find a conclusive theoretical argument that rules out the possibility of detecting such effects at this energy scale. 
\item{} The GIE experiment can be considered as the first instance of quantum computation that relies on gravity, which therefore goes beyond the standard quantum computation itself. It is an open question whether these effects could boost existing quantum computing techniques (rather than hindering them, as usually thought) by harnessing gravity in a post-quantum information scenario.
\end{itemize}

There is currently a race to implement the GIE experiment, and it has been claimed to be one of the most promising directions of research in fundamental physics, in recent years. If GIE were to be detected, plausible classical theories of gravity would be refuted. In particular, as we mentioned, it would be the first refutation of Einstein's general relativity, as a classical theory of gravity; and it would galvanise the field of quantum gravity, and confirm that a quantum theory of gravity is necessary to explain reality. If GIE were not to be detected, it would imply a number of equally interesting possibilities. It may be that gravity is classical after all; or that one of the principles (such as locality) taken for granted in analysing the experiment needs to be tweaked. It could also be that the quantum theory of gravity used to predict the effect is not correct, but gravity is still quantum, and a different method of quantisation is needed. It could also be that gravity is quantum, but there are other effects at play that are not predicted by current theories. All these possibilities are very exciting too, as they may require a radical change in the way we describe physical reality. It is promising that these and other avenues have been opened by exporting quantum information ideas to the quantum gravity scenario. We hope and expect this recently opened door will take us one step closer to the next revolution in physics. 

\maketitle

{\bf Acknowledgements} \;\; The authors are grateful to Sougato Bose, David Deutsch, Giuseppe Di Pietra, Tianfeng Feng, Marco Genovese, Gavin Moreley, Tomek Paterek and Fabrizio Piacentini for comments and criticism on earlier versions of this manuscripts. This publication was made possible through the support of the Eutopia Foundation, of the Gordon and Betty Moore Foundation, and of the ID 61466 grant from the John Templeton Foundation, as part of the The Quantum Information Structure of Spacetime (QISS) Project (qiss.fr). The opinions expressed in this publication are those of the authors and do not necessarily reflect the views of the John Templeton Foundation.

\section*{Appendix A: Quantum Network approximating GIE}

As explained in \cite{MAVE18, BHO}, it is possible to provide a quantum gate model that approximates  the quantum field theory Hamiltonian of the linear quantum gravity model. This is a consequence of the universality of quantum computation, \cite{BARENCO}, i.e., the property that certain sets of one and two qubit gates can be used to approximate any unitary evolution to arbitrarily high accuracy. Let us review the model briefly. The masses get entangled to the field, then the phases are generated through a generalised controlled-phase gate. Immediately after the action of the first beam splitter, the state of the two masses and the field is
$\ket{\phi_0}=\frac{1}{2}\sum_{a,b\in\{0,1\}}\ket{ab}\ket{\alpha}$,
where $\ket{\alpha}={\rm e}^{-\frac{1}{2}|\alpha|^2}\exp(\alpha (a^{\dagger}-a))\ket{0}$ is a coherent state representing the spatial modes of gravity - possibly a continuum. 

The two masses and the field then evolve into the state 
\begin{eqnarray}
\ket{\phi_{E1}}=U_1\ket{\phi_0}=\frac{1}{2}\sum_{a,b\in\{0,1\}}\ket{ab}\ket{\alpha_{a,b}}
\end{eqnarray}
where $U_1\doteq \sum_{a,b\in\{0,1\}}P_{ab}\otimes D(\xi_{ab})$ and $\ket{\alpha_{a,b}}=\ket{\alpha+i\sqrt{\xi_{ab}}}=D(\xi_{a,b})\ket{\alpha}\;$; we have defined the displacement operator as $D(\xi_{a,b})=\exp{(i\sqrt{\xi_{a,b}}(a^{\dagger}-a))}$ with $\sqrt{\xi_{ab}}$ being a real-numbered shift that depends on the coupling between the field and the masses, that brings about the desired phase-shift $\phi_{a,b}$ at the end (see below for more details). We have also defined the projectors $P_{ab}=P_a\otimes P_b$, with $P_{0,1}=\frac{(id\pm\sigma_z)}{2}$ being the projector operator for the location of each mass. We have also assumed that establishing the entanglement
between the field and the masses takes place on time-scales much faster than the process that transfers the phase $\xi_{a,b}$ back from the field to the masses, evolving their composite system to the state $\ket{\phi_{E2}}=U_2\ket{\phi_{E1}}\approx \frac{1}{2}\sum_{a,b\in\{0,1\}}\exp{(i\phi_{a,b})}\ket{ab}\ket{\alpha_{a,b}}$, where $U_2=\exp(w(a^{\dagger}a))$, $w$ is some real number with the property that $w\xi_{a,b}=\phi_{a,b}$, and, for the sake of this simple illustration, we have assumed to be in the regime where $|\alpha|$ is large and real (later the full linearised model will present the exact solution for any coherent state).  Finally, the interaction $U_1^{\dagger}$ between the field and the masses brings the field back to its original state and the masses remain entangled (to a degree depending on the phase): $\frac{1}{2}\sum_{a,b\in\{0,1\}}\exp{(i\phi_{a,b})}\ket{ab}\ket{\alpha}\;.$

In \cite{BHO} a simulation of the generation of GIE is discussed, and then implemented with NMR qubits; with a decoherence model it is also shown how the transition of the mediator from fully quantum to fully classical gradually hinders the entanglement generated between the two probes, as expected. 

A symmetric way of performing the maximally entangling gate between two qubits A and D via two other qubits B and C is represented by the circuit in Fig.~\ref{circuit1}.
First one applies a Bell gate between A and B, and between D and C; then one performs a controlled phase on qubits B and C; and finally one applies \CNOT\ gates between A and B, and D and C.
\begin{figure}
\includegraphics{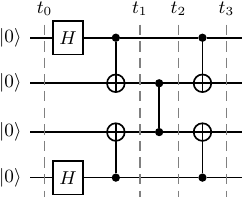}
\caption{{A symmetric quantum network generating entanglement between pairs of distant qubits. Following the quantum network notation, \cite{NIECHU}, each element represents a unitary interaction that is switched on for a finite period and then switched off. In particular, the box labelled as `H' denotes the Hadamard gate; the line with one full dot represents the Controlled-Not gate, and the line with two full dots the controlled Phase gate (each defined in the text).}}
\label{circuit1}
\end{figure}
The gates applied at their respective times are represented as follows:
\begin{equation}
\CNOT_{\alpha,\beta}= {\textstyle \frac{1}{2}} \left (id+q_{z\alpha}\right )+{\textstyle \frac{1}{2}} \left (id-q_{z\alpha}\right )q_{x\beta}
\end{equation}
\begin{equation}
\textsc{cph}_{\alpha,\beta}= {\textstyle \frac{1}{2}} \left (id+q_{z\alpha}\right )+{\textstyle \frac{1}{2}} \left (id-q_{z\alpha}\right )q_{z\beta}
\end{equation}
\begin{equation}
H_{\alpha}(t_n)= \frac{1}{\sqrt{2}} \left (q_{z\alpha}+q_{x\alpha}\right)\;.
\end{equation}

A fully optical implementation of this gate model is presented in \cite{SCIA}. 

\section*{Appendix B: Gravitational Decoherence Model}

We discuss a model demonstrating that the entanglement-induced decoherence of a massive superposition due to coupling to the quantized gravitational field in the linear regime. 
We pursue the analogy with the electromagnetic field and treat gravity in the linear regime in exactly the same way (see e.g. \cite{Boughn,Skagerstam}). By analogy, the gravity matter interaction Hamiltonian is given by:
\begin{equation}
H^G_{int} = \frac{1}{2} h_{\mu\nu} T^{\mu\nu} 
\end{equation}
where $T^{\mu\nu} \propto p_{\mu}p_{\nu}$
is the stress energy tensor of the single massive particle with momentum $p$ and $g_{\mu\nu} = \delta_{\mu\nu}+h_{\mu\nu}$. The quantized
gravitational field $h_{\mu\nu}$ is in the linear regime expanded as:
\begin{equation}
h_{\mu\nu} = \sum_\rho \int d^3k \{ a(k,s)\epsilon_{\mu\nu} (k,s) e^{ik_\lambda x^\lambda} + h.c.\}
\end{equation}
where $a(k,s)$ annihilates a graviton of momentum $k$ and spin $s$ and $\epsilon_{\mu\nu}$ is the graviton polarization. The notation $\phi = k_\lambda x^\lambda$ follows Einstein's summation convention and it represents the usual $kx-\omega t$ phase. We will work in the non-relativistic regime where only the $T_{00}$ component matters. This includes the component $mc^2$ as well as any other background potential (gravitational of otherwise) which contributes to the effective mass. This also implies that only the $h_{00}$ component of the field metric perturbation matters.

For our purposes we study a superposition of a massive object existing in two different locations which are therefore subjected to different local (gravitational or otherwise) fields.  Let us denote such a superposition, together with the state of the gravitational field, in the following manner:
\begin{equation}
(|\Psi_m (x)\rangle + |\Psi_m (x+\Delta x)\rangle)\otimes |\alpha\rangle 
\end{equation}
where $x$ is one location of the mass $m$ and $x+\Delta x$ the other one. Here $|\alpha\rangle$ is the initial state of the gravitational field, which would, in the static linear regime, simply be $\langle \alpha |\hat\Gamma |\alpha\rangle = \Gamma^i_{00} =-1/2h_{00,i}= \phi_G$. However, this is irrelevant as what matters are the dynamical degrees of freedom and the shift in the coherent state amplitude that they cause. This state subsequently evolves into
\begin{equation}
|\Psi_m (x)\rangle |\alpha + \delta \alpha \rangle + |\Psi_m (x+\Delta x)\rangle\otimes |\alpha-\delta \alpha \rangle \label{initial}
\end{equation}
where $\delta \alpha$ is the change of the coherent state due to coupling to gravity (to be given below). When the field is now traced out, the off-diagonal elements of the state of the particle are reduced by the amount $e^{-|\delta \alpha|^2}$. This is a time-dependent quantity and it represents an exponential amount of decoherence. Given that $\delta \alpha \propto \Delta E/E_P$ we find that the rate of decoherence to scale as $\gamma = |\delta \alpha|^2/t\propto (\Delta E/E_P)^2$. 

In order to obtain an estimate of the decoherence rate, we use the full linear regime Hamiltonian where the gravity-matter interaction Hamiltonian is:
\begin{equation}
H^{G}_{int} =\int_x dx \int_{k} dk \psi^{\dagger} (x)\psi (x) g_k (x) (a_{k\sigma}e^{i\omega_k t}+a^\dagger_{k\sigma}e^{-i\omega_k t})
\end{equation}
where, 
\begin{equation}
g_k = \frac{V(x) x^2\sqrt{16\pi G\hbar}}{4c^3\sqrt{V}}  \omega_k^{3/2} \; .
\end{equation}
We have ignored the $mc^2$ part of $T_{00}$ as it is the same for both states of the mass and will therefore not affect the relative phase between them. For simplicity, we have also surpressed the polarization degree of freedom and we have assumed that $x_ix_j=x^2$ since we are only interested in a rough estimate. Note that while the EM coupling is in the lowest order dipole, the gravitational coupling happens to be quadrapole and hence the scaling with $x^2$. Since the massive particle is assumed to be near stationary, the free part of the Hamiltonian is given by
\begin{equation}
H_{0} = \int dk (\hbar c k) a^\dagger_{k\sigma} a_{k\sigma}
\end{equation}
which is just the energy of the gravitational field. We will work in the rotating frame defined by $H_0$. Here we have that the annihilation and creation operators precess at the rate:
\begin{eqnarray} 
a_n & \rightarrow & a_n e^{i\omega_n t} \\
a_n^{\dagger} & \rightarrow & a_n^{\dagger} e^{-i\omega_n t} \; ,
\end{eqnarray}
and therefore the evolution due to the coupling becomes:
\begin{eqnarray}
U  =   \exp  \left\{  -i\int dx \psi^{\dagger}(x)\psi (x)\int_k dk g_k  \int_0^T(a_m e^{i\omega_m t} + a^{\dagger}_m e^{-i\omega_m t})dt \right\}
\end{eqnarray}
The evolution acts like a control gate, in the sense that it kicks the gravitational field only at the locations where the mass is present. Also, each momentum evolves independently of other momenta, and therefore, the relevant computation reduces to 
\begin{equation}
\exp \left\{-i g_k \int_0^T(a_k e^{i\omega_k t} + a^{\dagger}_k e^{-i\omega_k t})dt \right\}| \alpha_k \rangle \; ,
\end{equation}
where $T$ is the total time of evolution. This then allows us to calculate the shift of the gravitational coherent state and we readily obtain
\begin{equation}
| \alpha_k +\delta \alpha_k\rangle = |\alpha_k -i\frac{g_k}{\omega_k}(e^{i\omega_k T} -1)\rangle
\end{equation}
for each momentum state. The decoherence rate, when summed up over all momenta, then becomes: 
\begin{equation}
\frac{\gamma}{T} = \int_k dk g^2 (k) \rho (k) \frac{\sin^2 (\omega_k T) }{\omega^2_k} \; .
\end{equation}
The rate of decoherence depends on a number of assumptions, such as the relationship between the time of evolution and the relevant frequencies as well as the assumptions regarding the density of states $\rho (k)$ and the upper frequency cut-off $\Omega$ (see \cite{Ekert} for a detailed discussion). We can even include the effects of a finite temperature of the gravitational environment (all this requires in the above calculation is to assume that the field is in a thermal mixture of coherent states, as well as of course that we are in the usual Born-Markov regime). The final result is straightforward to anticipate using a simple dimensional analysis (which more detailed calculations corroborate as in \cite{Ekert}):
\begin{equation}
\gamma \approx \left(\frac{\Delta E}{E_P}\right)^2 \left( \frac{\Delta x}{c} \right)^n \frac{kT}{\hbar} \Omega^n
\end{equation} 
where $n$ is some positive power that depends on the above mentioned details, $\Delta E$ is the energetic difference between the two massive particle configurations, $E_P=m_Pc^2$ and $T$ is the temperature. Note that, if the upper frequency cutoff is given by $\Omega = \delta x/c$, we recover the result in \cite{Blencowe}.

\subsection*{Genuine Quantum Gravity Versus Classical Decoherence and Collapse}

What kind of rates are expected in practice given this formula? An electron superposed across two energy levels in an atom should be stable over a long period of time. On the other hand, a much more massive object, superposed across larger distances might decohere more rapidly.
Interestingly, if we assume that $\Delta E=mc^2$, the above rate at zero temperature
$\gamma = \left(\frac{\Delta E}{E_P}\right)^2 \left( \frac{\Delta x}{c} \right)^n \Omega^{n+1}$ reduces to the Penrose collapse formula. This is consistent with Penrose's interpretation that the fluctuations of the gravitational field produce an energy which, when divided by $\hbar$ gives the rate of the collapse. Here we see that there is a way of formally arguing for the gravitationally-induced collapse, but which is in no way different to any other form of quantum decoherence. Namely, imagine that the gravitational field is described by the quantised Christoffel symbols, $\hat \Gamma = \partial \hat g$. Then, the fluctuations in the gravitational field are \cite{Anandan}: 
\begin{equation}
\Delta^2 \Gamma = \int d^3 x \langle \alpha| \hat \Gamma^2 |\alpha\rangle - \langle \alpha| \hat \Gamma |\alpha\rangle^2
\end{equation}
Here gravitational decoherence is then seen as being induced by the fluctuations of the quantized gravitational field. This is the same as in the electromagnetic case of say the vacuum inducing spontaneous emission, or a more general dephasing and other such phenomena (see \cite{Ford} for the differences between the semi-classical and full quantum treatments). 

It is worth mentioning, however, that the above process does not involve dissipation. The energy of the massive superposition does not change, namely the diagonal elements are not affected by the decoherence. The linear quantum gravity model can also be used to obtain dissipation (say through spontaneous emission of gravitons). The standard calculation for the quadrupole emission gives us the rate:
\begin{equation}
\frac{dE}{dt} = -\frac{Gm^2a^4\omega^6}{c^5} =- \gamma_s \hbar \omega
\end{equation}
and this result can be also be obtained for say an electron making a transition within an atom by emitting a graviton in the linear regime \cite{Boughn,Weinberg,Dyson} (the formula is the same as the dipole emission of light, providing we make the substitutions $e\rightarrow \sqrt{G}m$ and $a\rightarrow a^2)$. The rate of emission $\gamma_s$ would be so small that the half life would be orders of magnitude bigger than the age of the Universe. As GIE shows, the fact that gravitons may not be observable does not imply that we cannot confirm the quantum nature of gravity.

Now we reach the crucial part of the discussion. Can we discriminate a genuine collapse from various possible entanglements with the quantum gravitational field? Yes, but only if we have access to other measurements of the field. If all we can do is measure the rate of decoherence of the superposed mass, then the origin of the decoherence cannot be established. What can be done is to discriminate an objective collapse from classical dephasing. Namely, imagine that the phases of the two states in the superposition evolve at different rates (as would be the case with when an atom is superposed at two different heights in Earth's gravitational field as in \cite{Brukner}). Then at some time $t$ the state would be 
\begin{equation}
e^{i\phi (x)t}|\Psi_m (x)\rangle + e^{i\phi (x+\Delta x)t} |\Psi_m (x+\Delta x)\rangle 
\end{equation}
Now, if there relative phase $\phi (x+\Delta x)-\phi (x)$ is large (compared to the time of measurement, say) then this will lead to the collapse of the interference between the two components, which is what we called classical decoherence. This decoherence, however, is only apparent and can actually be ``reversed". The standard technique is the spin echo, which just swaps the two elements of the superposition half way through the experiment. This equalises the two phases at the end, which therefore just become a global phase. Therefore the classical dephasing, unlike the objective collapse, can be undone. The key facts are that a genuine collapse can never be reversed (by definition), the classical dephasing can be reversed by acting on the system only and the decoherence due to entanglement can be reversed if we have access to and control of the environment causing the decoherence through coupling to the system.

\section*{Appendix C: Information-theoretic proof of the General Witness Theorem (GWT)}
This section presents a summary of the GWT proof within constructor theory of information, adapted from reference \cite{MAVE20}. {First we summarise informally the basics of constructor theory (CT) (see \cite{DEUMA} for the formal details). The fundamental concept in CT is that of a task. A task is the specification of a transformation expressed as a set of ordered pairs of input/output attributes. Attributes are sets of states of a given physical system. A physical system on which tasks can be performed we call `substrate'. If ${\bf a}$ and ${\bf b}$ are attributes, the attribute ${\bf (a,b)}$ of the composite system ${\bf S_1}\oplus {\bf S_2}$ is defined as the set of all states of the composite system where ${\bf S_1}$ has attribute ${\bf a}$ and ${\bf S_2}$ has attribute ${\bf b}$. I shall assume throughout the principle of locality ({\bf P1}), which requires that if a transformation operates only on substrate ${\bf S_1}$, then only the attribute ${\bf a}$ changes, not ${\bf b}$. It is well-know that this principle is satisfied by non-relativistic unitary quantum theory (see e.g. the discussion in \cite{DEUPA}), as well as by quantum field theories).  A {\sl variable} is a set of disjoint attributes. Given a task $T$, define its {\sl transpose} as the task obtained from $T$ by swapping each input attribute with the corresponding output attribute. A {\sl constructor} for a task $T$ is a system which whenever presented with the substrate of $T$ in any state belonging to one of the input attributes, it delivers it in one of the states of the allowed output attributes, and {\sl retains the ability to do that again}. A task is {\sl impossible} if the laws of physics impose a finite limitation on how accurately it can be performed by a constructor. Otherwise, the task is {\sl possible}. Note that a task being possible does not require there to be a perfect constructor in reality. Rather, the requirement is that the behaviour of such a constructor can be approximated to arbitrarily high accuracy, short of perfection, by a sequence of processes each of which can be realised in nature. If a task is possible, then for any given finite accuracy, it is possible to find a suitable environment which (under an allowed dynamical interaction) approximates a constructor for the task by delivering the substrates in the desired output attributes (within the specified accuracy). For the interested reader, in \cite{MAMA} there is a model of a possible task and approximate constructors in quantum theory, where it is highlighted that approximate constructors can be dissipative open systems. The constructor theory of information is based on an operational criterion expressing a requirement on possible tasks for systems that, informally, we consider capable of containing information. First one defines a class of substrates, {\sl information media}, by requiring that some tasks are possible on them - tasks that are conjectured to be sufficient for them to be capable of carrying information. In short, information media must have a variable $X$ with the property that it is possible to perform all the permutation tasks on $X$, and that it is possible to perform the task of copying all attributes in $X$ from one substrate to its replica. Any variable $X$ that can be copied and permuted in all possible ways is called an {\sl information variable}. A simple example of information medium is a qubit with an information variable being any set of two orthogonal states. Any two different information media (e.g. a neutron and a photon) must satisfy an {\sl interoperability principle}, which expresses elegantly the intuitive property that classical information must be copiable from one information medium to any other (having the same capacity), irrespective of their physical details, \cite{DEUMA}. More in detail, if $S_1$ and $S_2$ are information media, respectively with information variable $X_1$ and $X_2$, their composite system $S_1 \oplus S_2$ is an information medium with information variable $X_1\times X_2$, where $\times$ denotes the Cartesian product of sets.} 

Within the constructor-theoretic framework, we can restate the notion of non-classicality that the non-classicality witness is meant to assess, in constructor-theoretic terms. By a system being {\bf non-classical} one then means an information medium ${\bf M}$, with maximal information observable $T$, that has a variable $V$, disjoint from $T$ and with the same cardinality as $T$, with these properties:
\begin{enumerate}
\item{} There exists a superinformation medium ${\bf S_1}$ and a distinguishable variable $E=\{{\bf e_j} \}$ of the joint substrate ${\bf S_1}\oplus {\bf M}$, whose attributes ${\bf e_j}=\{(s_j, v_j)\}$ are sets of ordered pairs of states, where  $v_j$ is a state belonging to some attribute in V and $s_j$ is a state of ${\bf S_1}$;
\item{} The union of $V$ with $T$ is {\sl not} a distinguishable variable; \\
\item{} The task of distinguishing the variable $E=\{{\bf e_j} \}$ is possible by measuring incompatible observables of a {\sl composite superinformation medium} including ${\bf S_1}$, but impossible by measuring observables of ${\bf S_1}$ only.\\
\end{enumerate}
First, one considers three systems: $Q_A$ and $Q_B$ (two quantum probes), and $M$, the mediator, which has a `classical' observable $T$, and no other known degrees of freedom. In the case of gravity, the classical basis of the gravitational field would be the number operator, or its energy. Let A's descriptors be denoted by ${\bf q_A(t_0)}$, B's descriptors be denoted by ${\bf q_B(t_0)}$, and the mediator's descriptors be ${\bf c(t_0)}$. Let us assume that the observables (or measurable properties) of the three systems at time $t$, as well as their joint observables, are a fixed function of the triplet $({\bf q_A(t_0)}, {\bf c(t_0)},{\bf q_B(t_0)})$. In quantum theory, the function is given by the trace with the initial state's density operator, which never changes and therefore can be incorporated in the definition of the function. 
Assume that initially (at time $t=t_0$) the three systems are uncorrelated, for instance in a product state where a local observable of each subsystem is sharp with some value. So the mediator will have $T$ sharp with value say $t_0$. We assume that it is possible to run the experiment with two distinguishable initial conditions, say for instance: the case where the two masses are prepared in some state $S_{+}$ and in another distinguishable state $S_{-}$. 
Suppose that at time $t_3$ $Q_A$ and $Q_B$ end up in both cases being entangled.  This means that at time $t_3$ they have been prepared in one of two entangled states, call them $E_{+}$ and $E+{-}$, which are in turn distinguishable. Locality here can be used to conclude that in both configurations, the descriptors of $Q_A$, $M$ and $Q_B$ are an ordered triplet:
$$E_{\pm}\equiv \left({\bf q_A^{\pm}(t_2)}, {\bf c^{\pm}(t_2)},{\bf q^{\pm}_B(t_2)}\right)\;.$$
Now, we can use the assumption that $A$ and $B$ are not interacting directly, but the interaction is mediated by $M$. To a first approximation, this means that the interaction happens by letting first $A$ interact with the mediator $M$, at time $t_1$, and then $M$ interact with the qubit $Q_B$, at time $t_2$. 
Once more using locality, we find that at time $t_1$ the state of the three systems must be described by one of two triplets, according to whether the initial condition $S_+$ or the initial condition $S_-$ was used:

$$E_{\pm}\equiv \left({\bf q_A^{\pm}(t_1)}, {\bf c^{\pm}(t_1)},{\bf q^{\pm}_B(t_0)}\right)\;.$$
Locality is here used one more, because we have taken into account that the descriptor of system $B$ must not have changed since $t_0$, given that the interaction at $t_1$ only involves $A$ and $M$.
The proof proceeds by showing that the descriptors ${\bf c^{\pm}(t_1)}$ are: 1) disjoint (set-wise) from one another, and from the classical states of $M$; 2) that they are not distinguishable from the classical states of $M$; and 3) they are not distinguishable from one another; yet the joint state of $A$ and $M$ is (by means of measurements that involve both $A$ and $M$). 
These three properties make $M$ non-classical, in that it has a dynamical variable $V$, made of the two descriptors $\{{\bf c^{\pm}(t_1)}\}$ which is disjoint from the classical observable, and yet is not distinguishable from the latter, just like Heisenberg's uncertainty principle requires. 

It is very important to notice, as we have emphasised in the main text, that the non-classical variable $V$ of the mediator, unlike $T$, may not be an observable, in the sense that ${\bf c^{+}(t_1)}$ may not be distinguishable from ${\bf c^{-}(t_1)}$ in a single-shot manner; and that those two states may not be associated with a preparable state. {As we said in the main text, witnessing GIE does not allow one to conclude much about the properties of the states that gravity must be in while mediating the entanglement between the probes, in addition to the fact that they must have some degree of non-classicality.} 

\subsection{Towards a temporal version of the GWT}

{The gist of the argument presented in \cite{DIMA}, generalising the witness argument to the temporal domain, is as follows. Consider two systems: $Q$, a quantum system; and $M$, a system that we initially assume to have only one observable $Z_M$: the set of classical states is included in the span of its eigenstates, which forms a vector space. Assume now: (i) Conservation of a global observable on $M$ and $Q$, which (on $M$'s side) is a function of the `classical' variable $Z_M$ only ; (ii) The formalism of quantum theory. Assume also that $Q$ is a qubit, with $Z_Q$ representing its computational basis, and $M$ is a bit, so all their observables are two-fold.  
Let us now define the {\sl witnessing task} as the task where $Q$ evolves from an eigenstate of the conserved quantity $Z_Q$ to an eigenstate of another observable $X_Q$ that does not commute with $Z_Q$. This corresponds to creating quantum coherence on the system $Q$, in the basis where $Z_Q$ is diagonal.
We will now show that if the witnessing task can be achieved by letting $Q$ interact with $M$ only, then $M$ must be non-classical. We shall prove this result first by assuming an additive conservation law, and then a non-additive conservation law. 
Assumption (i) is formally expressed as the requirement that any allowed dynamical transformation $U_{MQ}$ acting on the joint system $M\oplus Q$ should satisfy:}  
\begin{equation}
[U_{MQ}, Z_M+Z_Q]=0 \label{CON}
\end{equation}
{We can readily see that this constraint together with the assumption that $M$ is a classical bit, with $Z_M$ being its only observable, leads to the impossibility of the witnessing task:  $Q$ cannot evolve if the conserved quantity is initially sharp. Note that the conservation law is enforced by requiring that the only allowed unitaries have the property that the operators $Z_M$ and $Z_Q$ are left unchanged. Hence, we cannot exploit $Z_M$ to ``read" the initial state of the probe $Q$ and therefore to induce the coherent evolution required for the witnessing task. Moreover, working with operators when enforcing the conservation law makes our argument independent of the initial state for the mediator $M$. This is an interesting point since we need not have direct control on the initial state of the system $M$ during the protocol we envisage.
For the only allowed unitary by \eqref{CON} must be a function of a Hamiltonian whose most general form is $\alpha Z_Q+\beta Z_M+\gamma Z_QZ_M$, with $\alpha, \beta, \gamma$ real constants. Such a unitary cannot make $Z_Q$ unsharp, if it is sharp initially. Hence if $Z_Q$ is made unsharp by interacting with $M$ only, and \eqref{CON} is satisfied, $M$ must be non-classical.}

In \cite{FEFE1} it is shown that it is possible to further relax the assumption of exact conservation law. This is in line with the fact that, if one assumes unitarity, the average conservation law and the exact conservation law of an additive quantity are equivalent \footnote{M. Ozawa, Private Communication.}.

\end{document}